\documentclass[12pt,preprint]{aastex}
\begin{document}

\parindent=1.0cm

\title{The Near-Infared Spectrum of the Nuclear Star Cluster: 
Looking Below the Tip of the Iceberg and Comparisons with Extragalactic 
Nuclei}

\author{T. J. Davidge}

\affil{Dominion Astrophysical Observatory,
\\Herzberg Astronomy \& Astrophysics Research Center,
\\National Research Council of Canada, 5071 West Saanich Road,
\\Victoria, BC Canada V9E 2E7\\tim.davidge@nrc.ca; tdavidge1450@gmail.com}

\altaffiltext{1}{Based on observations obtained at the Gemini Observatory, which
is operated by the Association of Universities for Research in Astronomy, Inc., 
under a cooperative agreement with the NSF on behalf of the Gemini partnership: the 
National Science Foundation (United States), the National Research Council (Canada), 
CONICYT (Chile), Minist\'{e}rio da Ci\^{e}ncia, Tecnologia e Inova\c{c}\~{a}o 
(Brazil) and Ministerio de Ciencia, Tecnolog\'{i}a e Innovaci\'{o}n Productiva 
(Argentina).}

\begin{abstract}

	Long-slit near-infrared (NIR) spectra of the Galactic nuclear star cluster 
(NSC) are discussed. The spectra sample the major axis of the NSC out to its half 
light radius. The absorption spectrum of the central regions of the NSC 
is averaged over angular scales of tens of arc seconds in order to sample globular 
cluster-like total luminosities, and the results are compared with model spectra. 
The equivalent widths of NaI$2.21\mu$m and CaI$2.26\mu$m outside of the center of the 
NSC, where light from nuclear bulge stars contributes a large fraction to 
the total flux, are consistent with solar chemical mixtures. In contrast, the 
equivalent widths of NaI$2.21\mu$m and CaI$2.26\mu$m near the center of the 
NSC are larger than expected from models with solar chemical mixtures, even after 
light from the brightest evolved stars is removed. The depths of 
spectroscopic features changing along the major axis of the NSC is consistent with 
imaging studies that have found evidence of population gradients in the NSC. 
That NaI$2.21\mu$m and CaI$2.26\mu$m are deeper than predicted for solar chemical 
mixtures over a range of evolutionary states is consistent with 
previous studies that find that the majority of stars near the center of the NSC 
formed from material that had non-solar chemical 
mixtures. The depths of the NaI$2.21\mu$m and CaI$2.26\mu$m features in the 
central regions of the NSC are comparable to those in the nuclear spectrum of the 
early-type Virgo disk galaxy NGC 4491, and are deeper than in the central 
spectra of NGC 253 and 7793. A spectrum of nebular 
emission and the youngest stars near the GC is also extracted. 
The equivalent widths of emission features in the extracted NIR spectrum 
are similar to those in the nuclear spectrum of NGC 253, and it is argued that this 
agreement is best achieved if the current episode of star formation near the center 
of the NSC has been in progress for at least a few Myr.

\end{abstract}

\section{INTRODUCTION}

	The Galactic center (GC) is a unique laboratory for 
examining the environment near a super-massive black hole (SMBH). 
With the development of speckle imaging and adaptive optics (AO) systems on 
large telescopes it has become possible to resolve individual stars 
within the central parsec of the Galaxy, enabling studies of 
their spectral energy distribution (SEDs; e.g. Eisenhauer et al. 2005; 
Buchholz et al. 2009) and motions (e.g. Eckart \& Genzel 1997; 
Ghez et al. 1998; Gillessen etal. 2009). The SEDs of these stars 
allow ages to be estimated via comparisons with isochrones, while 
their motions provide information about the central mass density. 
When combined, the spectrophotometric and positional information 
can be used to conduct tests of general relativity (e.g. Do et al. 2019). 
Studies of young stars near the GC also allow the mechanics of star 
formation in such an environment to be examined (e.g. Morris 1993; Ghez et al. 2003; 
Alexander 2005; Genzel et al. 2010).

	The innermost regions of the Galaxy are structurally 
distinct from the Galactic bulge. The area surrounding the GC is 
part of the nuclear nulge (NB), which has been described by Launhardt et al. (2002). 
The gas and stars that make up the NB define a flattened structure that 
extends over $\sim 230$ pc. The stellar content of the NB differs from that in the 
inner bulge, in that giants in the inner bulge appear to have uniformly 
old ages (Nogueras-Lara et al. 2018a), whereas some of the component 
stars of the NB are young.

	At the center of the NB is the nuclear star cluster (NSC), which is a 
flattened structure with a half light radius of 4.2 pc (Sch\"{o}del et al. 
2014). Feldmeier-Krause et al. (2017a) examine the structural properties of 
the NSC and find a mass of $2.1 \times 10^7$ M$_{\odot}$ within a 
8.4 parsec sphere. They also measure a M/L ratio of 0.9, which is consistent with 
that in other nuclear star clusters. There are kinematic signatures 
suggesting that the NSC has grown at least in part by the accretion of massive 
star clusters (Feldmeier et al. 2014). The NSC lacks a central cusp defined by 
bright stars (Do et al. 2009; 2013), although fainter objects may define a cusp 
(Yusef-Zadeh et al. 2012; Gallego-Cano et al. 2018; Sch\"{o}del et al. 2018). 
There is also a central concentration of ionized gas that is referred to as the GC 
mini-spiral (e.g. Ekers et al. 1983; Tsuboi et al. 2016).

	The NSC is a challenging observational target. Spectroscopic coverage is 
limited as heavy line-of-sight extinction effectively restricts studies to 
wavelengths longward of $\sim 1\mu$m, making traditional, well-calibrated 
spectroscopic probes of stellar content at visible wavelengths inaccesable. 
Crowding is an additional complication that 
limits the resolution of stars in the central parsec to $K \leq 
18$ (e.g. Davidge et al. 1997; Buchholz et al. 2009; Sch\"{o}del et al. 
2010, but see also Gallego-Cano et al. 2018). While the faint limit 
of current observations allows core helium-burning stars at the distance 
of the GC to be resolved, stars near the main sequence turn-off (MSTO) in 
color-magnitude diagrams (CMDs) of populations older than a few Gyr have yet to 
be sampled. Still, studies of resolved stars have revealed 
a broad range of ages with major recent star forming events within the 
past 10 Myr and at $10^8$ years (e.g. Krabbe et al. 1995). 
While much of the stellar content by mass along the NSC sight line in the NIR 
has an old or intermediate age, young stars with solar or higher masses 
have a total mass of 14000 to 37000 M$_{\odot}$ (Lu et al. 2013).

	Many of the recent studies that explore the stellar content near the center 
of the NSC have used AO-corrected IFU spectra, and almost all of these studies have 
examined the central parsec. A large population of early-type stars have been 
identified (Eisenhauer et al. 2005; Paumard et al 2006; Do et al. 2009; Bartko 
et al. 2009, 2010), some of which have circumstellar dust shells (Eckart et al. 2013).
The majority of early-type stars in the central arcsec are B main sequence stars, 
(Eisenhauer et al. 2005; Habibi et al. 2017). 
Young stars in the central parsec have a centrally-concentrated distribution 
(Paumard et al. 2006; Feldmeier-Krause et al. 2015), and outside of the central 
arcsec these define two disks that are kinematically distinct 
(Paumard et al. 2006; Bartko et al. 2009; 2010). The centrally-concentrated 
nature of the young stars is consistent with {\it in situ} formation 
(Feldmeier-Krause et al. 2015).

	Do et al. (2013) find that the luminosity function of late-type stars 
near the center of the NSC is like that of the bulge. In contrast, the 
mass function of recently formed stars near the center of the NSC 
is top-heavy (Paumard et al. 2006; Bartko et al. 
2010; Do et al. 2013; Lu et al. 2013). A top-heavy mass function can reconcile 
the observed X-ray emission with that expected from PMS stars (Nayakshin \& 
Sunyaev 2005; Lu et al. 2013). Still, Bartko et al. 
(2010) find that within 0.8 arcsec of SgrA* and at offsets in excess of 12 
arcsec ($\sim 0.5$ parsecs) young stars have a Salpeter mass function.

	The early-type stars near the center of the NSC are likely not 
coeval, although there are uncertainties assigning ages.
Paumard et al. (2006) and Bartko et al. (2010) suggest that 
the main population of bright early-type stars likely formed 6 Myr ago, 
while Lu et al. (2013) assign an age in the range 2.5 -- 5.8 Myr. 
IRS 7, which is the most luminous star near the GC, likely formed as part 
of a recent star-forming event, and Paumard et al. (2014) assign it an age of 6.5 
-- 10 Myr. There is also evidence for previous episodes of star formation during 
the past few tens of Myr as Bartko et al. (2010) find that the late-type B 
stars appear to be isotropically distributed, suggesting older ages. 
Star formation may also be occuring at the present day. Bartko et al. (2009) 
note that the shape of the disk and the mass distribution argues 
for {\it in situ} star formation, and Eckart et al. (2013) conclude 
that star formation near the center of the NSC is an ongoing process. 
Pei$\beta$ker et al. (2020) discuss the properties of moderately faint (K $\sim 18$) 
sources that have a thermal SED, and are identified as possible PMS objects 
close to SgrA*. That star formation is on-going is consistent with the presence of 
gas flows and isolated dust structures near SgrA* (e.g. Paumard et al. 2004).

	The current wealth of imaging and spectroscopic studies have only sampled 
the tip of the stellar content iceberg. When working at their diffraction 
limits, the next generation of very large telescopes will 
resolve stars as faint as K $\sim 22$ near the GC (e.g. Skidmore et al. 2015), 
and so the long-term prognosis for resolving the upper 
portions of the main sequence in old populations near the center of the NSC 
is positive. Still, achieving this goal is many years in the future. 
Even with the expected gains in angular resolution only a fraction 
of the stellar content will be examined, and it will not 
be possible to resolve main sequence stars with sub-solar masses within a 
few arcsec of SgrA*. 

	Studies of integrated light provide a means of investigating the 
properties of all stars in a system, including the faint main sequence stars that 
define the mass function among low mass stars and fainter pre-main 
sequence objects, albeit in an encoded format. 
The problems associated with binarity in -- say -- counting 
stars to assess the mass function or interpreting CMDs 
is removed when examining integrated light, as there is an unbiased 
and complete sampling of all objects. The composite spectrum of 
the GC may also yield clues into the origins of the stars 
in the NSC through an examination of chemical 
mixtures; for example, did stars in the NSC form from gas 
that migrated inwards from the Galactic thin disk, or in dense 
massive star clusters that subsequently spiralled into the central regions 
of the Galaxy (e.g. Antonini et al. 2012; Arca-Sedda \& Cappuzzo-Dolcetta 2014)? 
SEDS constructed from integrated spectra can also examine the 
influence that the GC environment has on stellar properties, testing the suggestion 
that some stars near SgrA* have been stripped of their envelopes via stellar 
collisions (e.g. Alexander 1999; Genzel et al. 2003; Dale et al. 2009) or 
by interactions with dense proto-stellar clumps (Amaro-Seoane \& Chen 2014), as this 
will result in a deficiency of spectroscopic signatures from cool evolved stars.

	Studies of the integrated NIR spectrum of the NSC 
provide a direct means of assessing the properties of its component stars over 
a range of luminosities, while also identifying similarities between the GC,
nearby galaxy nuclei, and star-forming regions in general. 
Integrated spectra provide information that is complementary to that obtained 
from the imaging surveys of the inner Galaxy discussed earlier in 
this section. While individual bright stars can be resolved 
in the central parsec of the Galaxy, this is not yet possible for even the nearest 
galaxies. Spectra of the NSC thus provide a bridge that links the information 
obtained for the inner Galaxy to the nuclei of more distant systems, 
and provide a means of answering questions such as: Do the properties 
of the stars and star-forming regions near the center of the NSC differ from those 
in the rest of the Galaxy? and in other galaxies? 
Is the duty cycle of star-forming activity near the GC consistent 
with that in other galaxy nuclei? Is there evidence for ionization from 
sources other than the photospheres of hot stars?

	Figer et al. (2000) discuss the integrated spectroscopic 
characteristics of the area within a few arcsec from SgrA*. Using spectra 
that span 2 to $2.4\mu$m, they find that light within 
a few tenths of an arcsec of SgrA* has characteristics that are 
consistent with those of early-type stars. 
Spectroscopic signatures of late-type stars, such as the first overtone 
CO bands, only become prominent at radii in excess of an arcsec from SgrA*. 
However, the angular coverage of these data is very small.

	Shields et al. (2007) discuss observations of emission line nuclei that 
were obtained for the Survey of Nearby Nuclei with STIS (SUNNS), and examine the GC 
in the context of those objects. As extinction makes recording the visible wavelength 
spectrum of the GC problematic, they estimate the H$\alpha$ luminosity of the 
GC from free-free emission measurements, and conclude that it is comparable to 
that in the nuclear regions of most galaxies in the SUNNS sample. Furthermore, 
the absence of a strong compact X-ray source and the low-ionization state 
of lines suggests that the center of the NSC is neither a Seyfert nucleus nor a 
LINER. Rather, it appears to be an HII nucleus, or possibly a transition object.

	There have been more recent studies of the inner Galaxy that 
use spectra of integrated light. Feldmeier et al. (2014) obtained integrated light 
spectra of a $4 \times 3.5$ arcmin area that includes SgrA*. This sky coverage was 
obtained using drift scanning techniques, and the angular resolution 
perpendicular to the slit is $\sim 2$ arcsec. The 
primary scientific driver for these data was to examine kinematics, 
and the wavelength coverage was restricted to $2.29 - 2.41\mu$m, thereby sampling 
the three first overtone CO band heads. Feldmeier-Krause 
et al. (2015) obtained KMOS spectra in the $K-$band spectra of a $65 
\times 43$ arcsec area that includes SgrA*. These data were used to examine the 
spectra of individual stars and also map the angular distribution of 
Br$\gamma$, HeI, and H$_2$ 1--0 S(1) emission lines.

	More direct comparisons with external nuclei have been made in the 
mid-infrared (MIR). Simpson et al. (1999) examine 
the strengths of MIR emission lines in central Galactic fields and find that 
they are similar to those in a low excitation HII region rather than an 
AGN. They note similarities with M82, pointing out that the diffuse radiation 
field inferred from the line strengths are indicative of an aging starburst.

	In the present paper long-slit spectra 
in the wavelength interval $0.9 - 2.4\mu$m 
that sample the central few parsecs of the NSC along its major axis 
are discussed. The spectra were recorded with the Flamingos-2 (F2) 
imaging spectrograph on Gemini South (GS). 
The use of a long slit permits coverage of parts of the NSC 
that are well outside of the central parsec, which has been the 
target of most previous spectroscopic studies. A wider range of wavelengths 
are also covered. The inclusion of spectra in the $J-$band are of particular use for 
identifying bright foreground disk stars that will otherwise contaminate 
the integrated spectrum. The NIR spectrophotometric properties of the 
NSC in four different angular intervals are compared with model spectra to gain 
insights into the chemical mixture of stars that span a range of evolutionary states, 
and assess the stellar and chemical homogeneity of the NSC.
Comparisons are also made with other galactic nuclei. 

	Details of the observations and their reduction are discussed in Section 2. 
Basic characteristics of the spectra are examined in Section 3, while 
comparisons with model spectra are made in Section 4. 
Comparisons are made with spectra of galaxy nuclei in Section 5, where a 
spectrum of the young component near the GC is also extracted. A 
summary and discussion of the results follows in Section 6.

\section{OBSERVATIONS \& REDUCTIONS}

	The spectra were recorded with F2 (Eikenberry et al. 2004) 
on GS for program GS-2018A-Q-414 (PI: Davidge). The F2 detector is 
a Teledyne Hawaii-2 $2048 \times 2048$ array, and each pixel subtends 
0.18 arcsec along the spatial direction. The dispersing elements are grisms. 

	There are challenges associated with obtaining a spectrum of the 
NSC that can be compared with model spectra and the nuclear spectra of other 
galaxies. Representative stellar content must be sampled to avoid stochastic 
effects. This is of particular relevance in the NIR, as much of the NIR light from 
systems that are older than a few Myr comes from intrinsically 
luminous highly evolved stars that have short evolutionary timescales. Such 
objects are susceptible to 
stochastic sampling effects, even in dense environments. Some of the stellar 
content near the center of the NSC also has a complicated projected distribution, 
with the youngest stars located in warped disks (e.g. Bartko et al. 
2009; 2010). These sampling issues can be mitigated by observing 
moderately large areas at the expense of angular resolution. For this study 
spectra are constructed from areas that cover 50 arcsec$^2$ in the central 
regions of the NSC, and 65 arcsec$^2$ in the outer regions of the NSC. 

	The spectra were recorded with the JH and HK 
grisms, and so deliver a combined wavelength coverage 
of $0.9 - 2.4\mu$m. The slit used for these observations has a 
length of 4.3 arcmin and a width of 0.4 arcsec. The spectral resolution with 
this slit width varies between $\frac{\lambda}{\Delta \lambda} =$ 500 and 
1100, depending on wavelength. The NSC has a half light 
radius of 4.2 pc (Sch\"{o}del et al. 2014), which corresponds to $\sim 110$ 
arcsec at the distance of the GC; the length of the F2 slit 
thus exceeds the half light diameter of the nuclear star cluster.

	The slit was oriented to parallel the Galactic Plane (GP) for all 
observations. Two slit pointings centered on SgrA 
that are offset $\pm 1$ arcsec perpendicular to the GP were 
observed to increase spatial coverage. The slit 
positions sample the areas of high Br$\gamma$ and HeI$2.06\mu$m 
emission in the central regions of the NSC, as shown in Figures 7 and 8 of 
Feldmeier-Krause et al. (2015).

	In order to remove the telluric and thermal emission 
that dominates the sky signal at these wavelengths it is desireable 
to sample areas of pristine sky. This is a challenge as the NSC is located in 
the Galactic bulge, and so observing completely blank sky fields requires pointing 
offsets of tens of degrees, thereby complicating efforts to obtain a 
reliable local background measurement. Fortunately, there 
are extended areas of very high extinction near the NSC that more-or-less 
monitor the background sky. Spectra of a high-opacity dust lane that parallels 
the GP were thus recorded. The dust lane is $\sim 1$ arcmin to the south east of 
the GP. The locations of the science and background slits are shown in Figure 1. 

\begin{figure}
\figurenum{1}
\epsscale{1.0}
\plotone{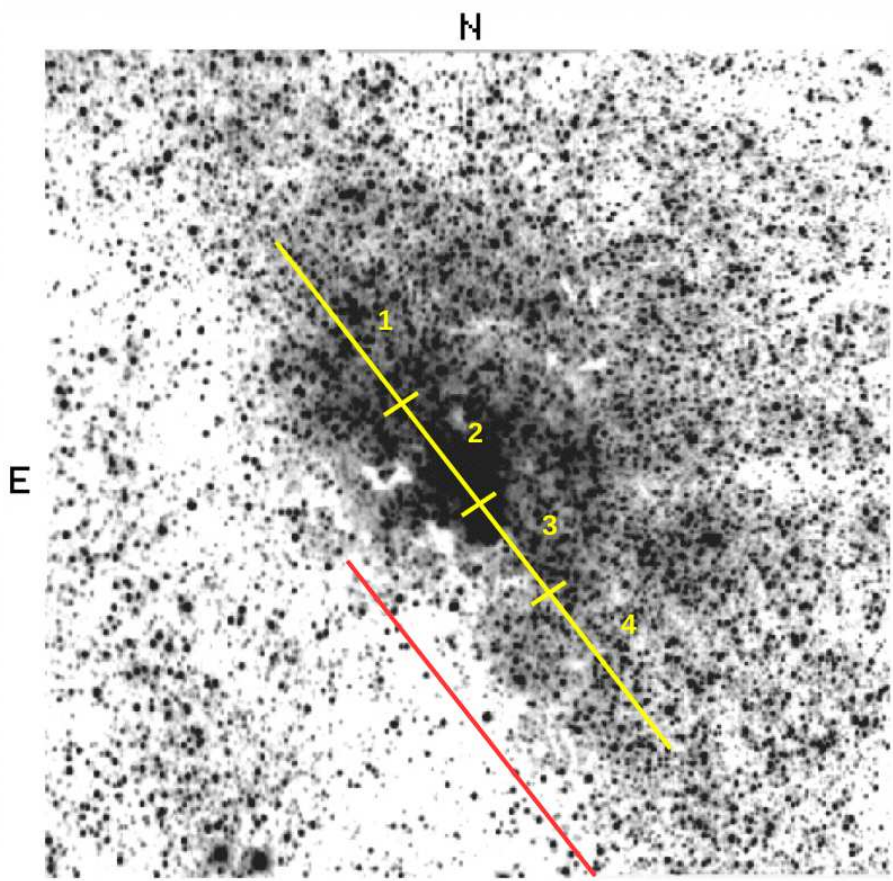}
\caption{Slit locations. The background image is a $6.2 \times 6.2$ arcmin 
$K$ image centered on SgrA from Figure 1 of Davidge (1998). The NSC (yellow line) 
and background (red line) slit locations are indicated. While only one yellow line 
is shown, there are actually two NSC slit pointings that are offset by $\pm 1$ arcsec 
perpendicular to the Galactic Plane. The background slit extends off 
of the image. The yellow hash marks demarcate the regions that are defined in 
Section 3, and these sample areas of 50 arcsec$^2$ for each of Region 2 and 3, 
and 65 arcsec$^2$ for each of Region 1 and 4. Regions 2 and 3 sample the densest 
regions of the GC mini-spiral and the center of the nuclear star cluster.}
\end{figure}

	Pockets of high extinction appear as lighter areas in Figure 1, 
and there are clear angular variations in the 
extinction in the direction of the NSC. The largest variations in extinction tend to 
occur perpendicular to the GP, with the areas of 
highest extinction being found at southern Galactic latitudes, which 
is the region where the background slit measurements are made. The positioning of 
the F2 slit was checked for large-scale extinction variations. While portions of the 
slit to the lower right of the GC in Figure 1 skirt areas of high extinction, 
the slit avoids these. While there are comparatively small areas along 
the slit where the signal dips to levels that are consistent with 
high levels of extinction, these do not dominate the overall signal. 
The impact that these small pockets of extinction might have on the 
spectrum are mitigated by observing two slit positions that are offset 
perpendicular to the GP.

	Exposures of the NSC and the background dust lane were alternated in 
the sequence NSC -- dust lane -- NSC -- dust lane etc. The NSC and dust lane 
observations were dithered by a few arcsec along the slit from exposure-to-exposure 
to mitigate against light from individual stars and cosmetic defects on the 
detector. In addition to the NSC observations, spectra of 
early-type stars were usually recorded at the beginning and 
end of a nightly observing session to monitor telluric absorption features. These 
were accompanied by exposures of a continuum and Ar emission arc lamps 
that are located in the Gemini calibration unit to provide flat-field and 
wavelength calibration information. Spectra were recorded on four nights, 
and a summary of these observations can be found in Table 1. 

\begin{table*}
\begin{center}
\begin{tabular}{lcl}
\tableline\tableline
Date & Grism & n$_{exp}$\tablenotemark{a} \\
 (UT; 2018) & & \\
\hline
July 23 & JH & $15 \times 120$sec \\
July 25 & HK & $57 \times 15$sec \\
August 1 & JH & $14 \times 120$sec \\
August 4 & JH & $14 \times 120$sec \\
\tableline
\end{tabular}
\end{center}
\caption{Summary of Observations}
\tablenotetext{a}{Number of SgrA exposures. A similar number of exposures were 
recorded of the background field (Section 2).}
\end{table*}

	Instrumental and atmospheric signatures were removed 
from the spectra by applying a standard recipe 
for the reduction of NIR spectra. The basic steps are: (1) the 
subtraction of background spectra from the NSC spectra, (2) division by flat-field 
exposures, (3) geometric rectification to remove distortions that are introduced 
by the F2 optics, (4) wavelength calibration, and (5) division by the spectrum of a 
telluric calibration star. The processed spectra for each slit position were aligned 
to correct for dither offsets along the slit and then averaged together.
The spectra from both NSC slit positions were then combined.

	The spectra were normalized to a pseudo-continuum that was identified by 
fitting a low-order Legendre polynomial. An iterative rejection algorithm was 
applied to suppress emission and absorption features when 
identifying the pseudo-continuum. A low-order fit reduces the risk of artificially 
flattening broad molecular features, and can track the pseudo-continuum 
since the division by the telluric standard star (step 5, above) corrects for 
high order wavelength variations in optical response due to atmospheric transmission 
and the observing system (i.e. telescope and instrument optics). 

\section{CHARACTERISTICS OF THE SPECTRA}

\subsection{IRS 7 and Systematic Effects}

	The red supergiant (RSG) IRS 7 is the most luminous star near the GC, 
and is sampled with the F2 spectra. IRS 7 has been well-studied, and so 
serves as a benchmark to assess systematic effects in the F2 spectra 
that might arise due to the observational set-up (e.g. spectral resolution) 
and subsequent processing (e.g. placement of the pseudo-continuum). 
Carr et al. (2000) find T$_{eff} = 3500 \pm 50$K 
for IRS 7, and assign it a near-solar [Fe/H], although they also find non-solar 
CNO abundances that they attribute to deep mixing. Paumard et al. (2014) 
assign IRS 7 an effective temperature of $3600 \pm 195$ K. The effective temperature 
of IRS 7 thus agrees with that of the RSG $\alpha$ Ori (T$_{eff} = 3600 
\pm 66$; Haubois et al. 2009), and Carr et al. (2000) compare the IRS 7 
spectrum with that of $\alpha$ Ori. Carr et al. (2000) conclude that [Fe/H] 
in $\alpha$ Ori is similar to that of IRS 7, but also find that the first overtone 
CO bands in IRS 7 are weaker than those in $\alpha$ Ori.

	We compare the spectrum of IRS 7 extracted from the F2 observations 
with that of $\alpha$ Ori to assess possible systematic 
effects in the F2 spectrum. The 
spectrum of $\alpha$ Ori recorded by Rayner et al. (2009) 
for the IRTF library was downloaded from the archive website, 
\footnote[1]{http://irtfweb.ifa.hawaii.edu/~spex/IRTF\_Spectral\_Library/} 
smoothed, and re-sampled to match the spectral resolution and wavelength sampling 
of the IRS 7 spectrum. The result was normalized to a pseudo-continuum 
using the same procedure that was applied to all F2 spectra. The processed $\alpha$ 
Ori spectrum is compared with the IRS7 spectrum in Figure 2.

\begin{figure}
\figurenum{2}
\epsscale{0.8}
\plotone{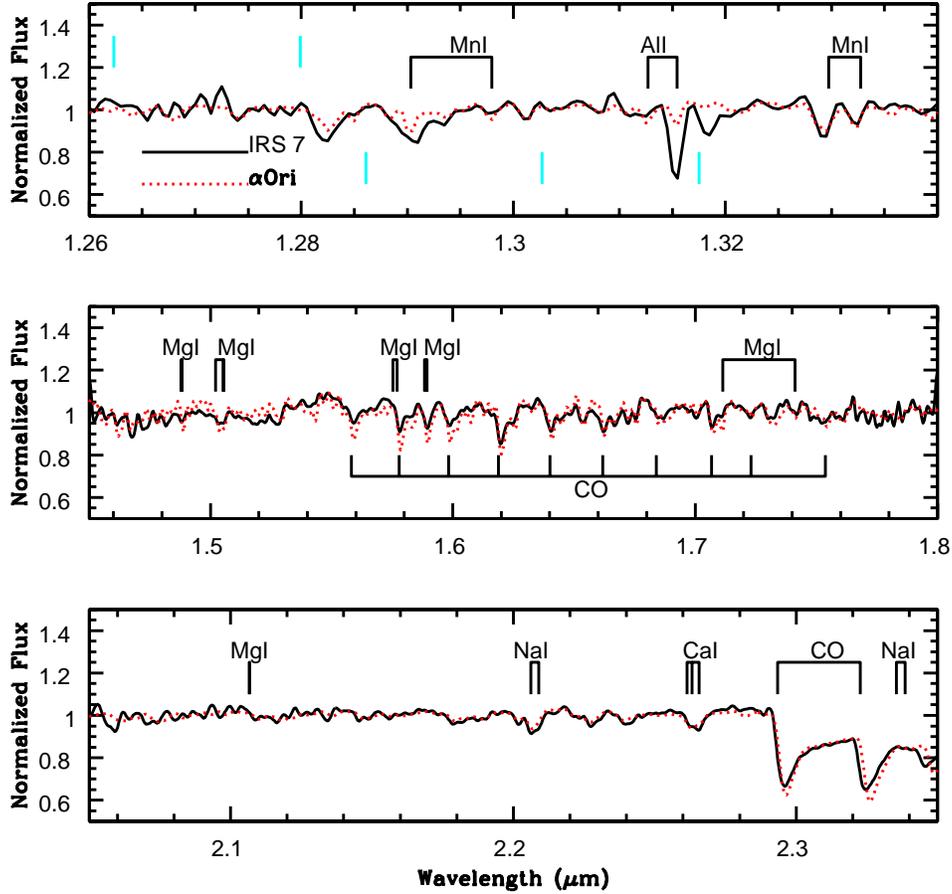}
\caption{Spectra of IRS 7 (black line) and $\alpha$ Ori (red 
dotted line). The IRS 7 spectra were extracted from the F2 
observations, while the $\alpha$ Ori spectra were constructed from 
a spectrum in the Rayner et al. (2009) library that was processed to match the 
spectral resolution and wavelength sampling of the IRS 7 spectra. The spectra of 
both stars have been normalized to a pseudo-continuum. The locations of 
diffuse interstellar bands (DIBs) in the $J-$band are marked with cyan lines in 
the top panel, and there is no obvious correspondence with features in the IRS 7 
spectrum. Atomic and molecular features are identified 
using the rest frame vacuum wavelengths listed in Tables 
7 and 10 of Rayner et al. (2009). There is good agreement between 
the strengths of most atomic features at wavelengths between $1.3\mu$m and 
$2.3\mu$m; the agreement at wavelengths shortward of $1.3\mu$m degrades due to the 
lower S/N ratio of the IRS 7 spectrum at these wavelengths. 
The depths of the first- and second-overtone CO bands are shallower in the IRS 7 
spectrum, in agreement with the relative strengths of the CO bands in the 
spectra of these two stars at higher spectral resolutions (Carr et al. 2000). 
Note that the depths of the NaI$2.21\mu$m and CaI$2.26\mu$m features in 
the IRS 7 spectrum match those in the $\alpha$ Ori spectrum at the 1 -- 2\% level.} 
\end{figure}

	The agreement between the IRS 7 and $\alpha$ Ori spectra near 
the $1.33\mu$m MnI lines in the top panel of Figure 2 is excellent. However, the 
agreement between the two spectra degrades at wavelengths $< 1.3\mu$m. This is 
due to the lower signal level in the IRS 7 spectrum at these wavelengths, 
compounded by the challenges removing telluric emission lines from faint sources. 

	IRS 7 is heavily reddened (E(B--V) $\sim 8$), 
raising the possibility of contamination from diffuse interstellar 
bands (DIBs), which could complicate comparisons with $\alpha$ Ori. 
The wavelengths of DIBs in the $J-$band are indicated 
in the top panel of Figure 2. There is poor agreement between absorption features 
in the IRS 7 spectrum and the wavelengths of DIBs, although the 
DIB near $1.32\mu$m is a possible exception. Thus, DIBs are not 
responsible for the differences between the IRS 7 and $\alpha$ Ori $J-$band spectra.

	Moving to the middle and bottom panels, 
the first and second overtone CO bands in the IRS 7 spectrum are clearly 
weaker than those in $\alpha$ Ori, in agreement with what was found by 
Carr et al. (2000). This is noteworthy since the deep nature and broad wavelength 
coverage of the first overtone CO bands cause these features to overlap at 
the spectral resolution of the F2 observations, making them susceptible to 
possible uncertainties in the placement of the pseudo-continuum. In contrast 
to the CO bands, the depths and widths of most atomic features in the two spectra 
more-or-less match at wavelengths between 1.5 and $2.4\mu$m, and this includes 
the NaI$2.21\mu$m doublet and the CaI$2.26\mu$m triplet. Neither of these features 
are resolved into their component lines at the wavelength resolution of the F2 
spectra.

	The depths of the NaI$2.21\mu$m and CaI$2.26\mu$m features 
in IRS 7 and $\alpha$ Ori in Figure 2 agree to within 1 - 2 percent. 
This level of agreement is of interest as Sellgren et al. 
(1987), Terndrup et al. (1991), and Blum et al. (1996) found that NaI$2.21\mu$m and 
CaI$2.26\mu$m tend to be deeper in stars near the GC than in the solar neighborhood, 
and this includes IRS 7. Using high resolution spectra, Carr et al. (2000) 
find that the Na lines in the $2.21\mu$m Na doublet in IRS 7 and $\alpha$ Ori are 
similar in strength. However, the depths of Sc, V, and CN lines that are at 
wavelengths similar to the NaI doublet differ from those in $\alpha$ Ori, 
in the sense of being deeper in IRS 7. While the spectral 
resolution of the F2 spectra in the $K-$band is higher than in the 
Blum et al. (1996) spectra ($\frac{\lambda}{\Delta \lambda} =650$ vs. 570), this 
is not sufficient to resolve individual lines near NaI$2.21\mu$m
(e.g. Figure 6 of Carr et al. 2000). 

	The placement of the pseudo-continuum may be responsible for the apparent 
discrepancy between the relative strengths of the NaI$2.21\mu$m and CaI$2.26\mu$m 
measurements in IRS 7 and $\alpha$ Ori in the F2 spectra when compared 
with past studies. The wavelength region that contains 
the NaI and CaI lines coincides with R$_2$ bands of CN. 
The F2 spectra have been processed by applying an 
unbiased and uniform algorithm for identifying continuum location that 
is based on broad wavelength coverage and is not sensitive to molecular 
absorption features or residuals from telluric emission/absorption features. In 
fact, the continuum levels in the IRS 7 and $\alpha$ Ori spectra in the vicinity of 
the NaI$2.21\mu$m and CaI$2.26\mu$m lines in Figure 2 are in excellent agreement. 
Later in the paper it is shown that the NaI$2.21\mu$m and CaI$2.26\mu$m features in 
the integrated NSC spectra (1) vary with position along the F2 slit, and (2) are 
deeper than in models with a solar chemical mixture. Therefore, given the similar 
depths of NaI$2.21\mu$m and CaI$2.26\mu$m absorption 
in the IRS 7 and $\alpha$ Ori spectra in Figure 2, 
if the F2 spectra systematically underestimate the depths of these features then 
the differences in line strenghs that are discussed later in the paper are 
likely lower limits to actual differences.

\subsection{Extraction of Spectra and Assessment of Stochastic Effects} 

	Co-added spectra in four angular intervals have been considered to examine 
variations in the stellar content of the NSC along the slit, and 
the boundaries of these intervals are indicated in Figure 1. Regions 1 and 4 
probe the outer regions of the NSC, while Regions 2 and 3 probe the 
center of the NSC and the GC Mini-spiral, which contains the youngest stars near 
the center of the NSC. While the use of wide angular gathers blurs spatial 
resolution, it boosts the S/N ratio while also suppressing 
the influence that individual bright stars might have on the spectra. 

	Each interval contains bright stars for which individual spectra can be 
traced. Some have blue SEDs, and stand out with respect to 
the majority of individual sources due to their high 
S/N ratios in the $J-$band. These are foreground stars, and light from 
these objects was removed prior to combining spectra in each region. 
Foreground star light was removed over the entire wavelength range 
in the angular interval where the PSF exceeded that of the surrounding 
background. The extraction area was filled by interpolating between the 
extraction limits.

	The majority of individual stars in the F2 spectra 
have distinctive red SEDs. These are heavily reddened 
objects beyond the foreground Galactic disk, and the majority of 
these objects are likely associated with the NSC, the NB disk, and the bulge.
While much of the obscuration is in the foreground disk, there is also a dust 
component associated with SgrA (Davidge 1998). Still, the foreground disk 
extinction is more than sufficient to produce objects with highly reddened SEDs.

	Individual bright stars can contribute significantly to the integrated 
light. IRS 7 is a prime example, and light from such objects can 
affect the integrated spectra. Two sets of spectra were thus constructed for 
each region. One includes all signal (but not that 
from foreground stars -- see above) in each region. The other
excludes light from all stars that have $K < 12$. Bright 
NSC stars were identified and their light removed using the same procedure 
described in the previous paragraph for foreground stars.

	The $K=12$ threshold corresponds to a peak signal that departs from that of 
the unresolved body of light in the $H$ and $K$ spectra of Regions 2 and 3 
at the $2\sigma$ level. A fainter cut-off would have been defined 
in Regions 1 and 4, where the local background is substantially 
lower, and/or if the signal had been restricted to the $K-$band instead of 
both $H$ and $K$. The RGB-tip occurs near M$_K \sim -6.6$ for systems 
with near-solar metallicities (Ferraro et al. 
2000), which corresponds to $K \sim 10$ in the NSC. Therefore, the stars that 
are removed are not exclusively luminous young objects associated with recent star 
formation; rather, many are first or second ascent giants that likely formed 
Gyr in the past.

	The seeing was $\sim 1$ arcsec FWHM when these data 
were recorded, and so light from bright stars that are outside of the 
0.4 arcsec slit will be present. This will occur most frequently in 
Regions 2 and 3, where crowding is most severe. Bright 
sources were identified solely from their appearance 
as extended objects in the slit spectra, and so 
light from very bright stars that are within 1 -- 2 arcsec of the 
slit will still be identified, albeit with an amplitude that does not correspond 
to their actual brightness. Any light from bright stars that 
is missed is then at a level that is comparable to 
that from the main body of unresolved stars.

	There are indications that contamination from bright stars that are 
within a few arcsec of the slit is not significant. Light that is 
dominated by a single bright star will have a lower velocity dispersion than 
that in the main body of stars. In fact, velocity dispersions measured 
in each region are found to increase when the light from bright stars is removed, as 
expected if the light is dominated by unresolved sources. In addition, 
the spectroscopic properties of each region changes when light from 
bright stars is removed, in the sense of features weakening after the 
removal of light from bright stars. This is most obvious in the CO(2,0) indices (see 
below), and is consistent with the successful removal of light from 
bright, late-type stars.

	We conclude that while there might be residual contamination from the 
outer PSF wings of bright stars, it is at a level that is likely small when compared 
with the removal of light from stars that are detected on or close to the 
slit. Given that some contamination is present then 
the differences found here that are attributed to the removal of bright stars 
are likely lower limits. Spectra obtained with a higher angular 
resolution, where the contamination from light produced by bright stars 
off of the slit will be reduced, should then find even larger differences 
than those discussed below.

	Region 2 includes IRS 7, which is by far the most luminous object 
near the GC ($K \sim 7$, M$_K \sim -9.5$), and is a star that contributes 
substantially to the total light from Region 2 in all three passbands.
Given the brightness of IRS 7, the influence of bright 
stars on the integrated spectrum of Region 2 sets an upper limit to what is 
seen in the other regions, and so the Region 2 spectrum is examined in detail to 
provide a conservative means of assessing how light from bright stars affects 
the final spectra. The result of removing light from bright stars on the integrated 
spectra of Region 2 is explored in Figure 3, where spectra of Region 2 with and 
without light from resolved stars are compared. 

\begin{figure}
\figurenum{3}
\epsscale{1.0}
\plotone{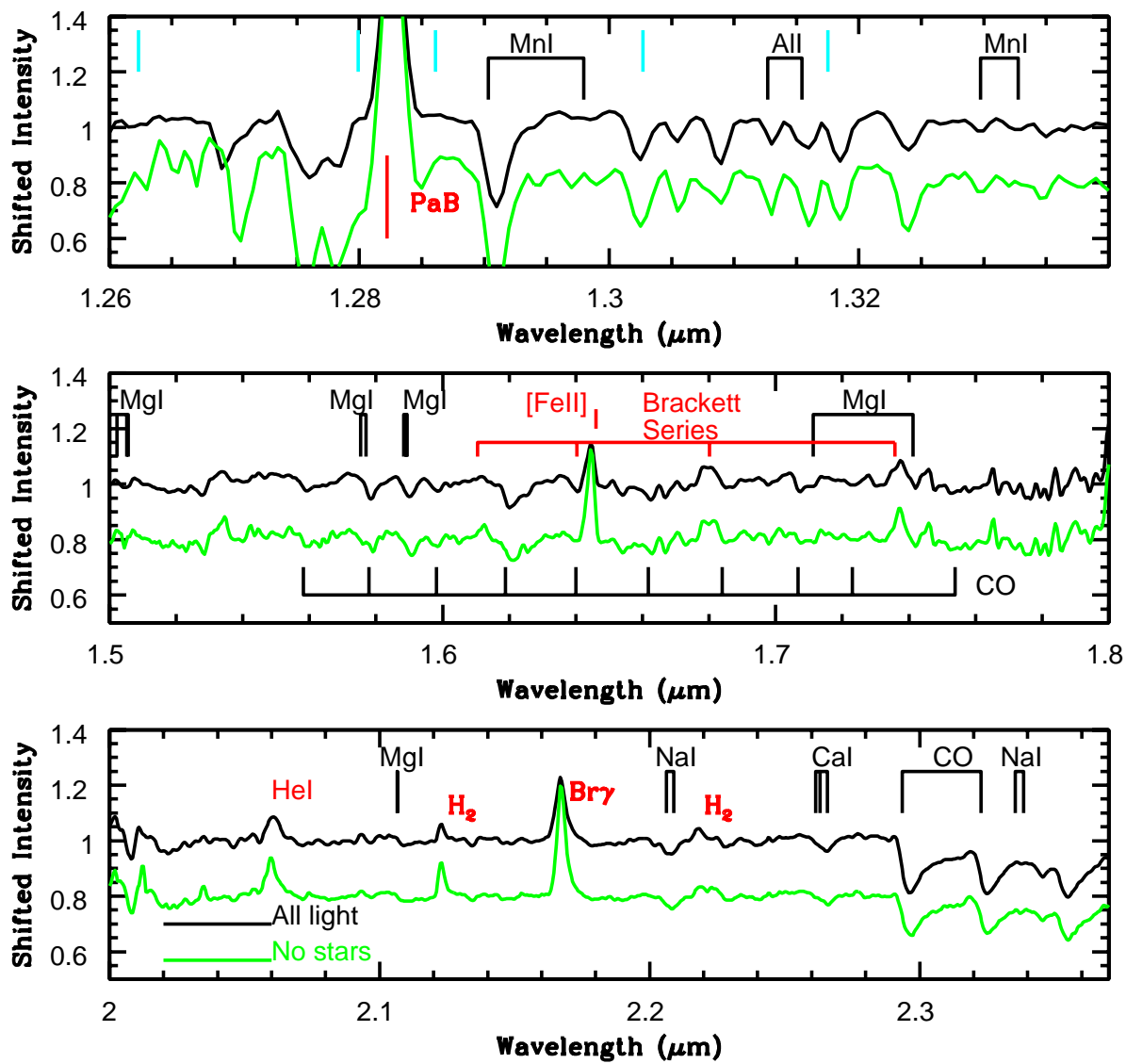}
\caption{Assessing the influence of bright stars on the spectrum of Region 2. The 
black lines are spectra in which light from resolved bright stars is included, 
while the green lines are spectra where the light from bright sources has been 
removed. The spectra have been offset vertically for display purposes. 
Cyan lines in the top panel mark the expected locations of DIBs. 
The removal of light from bright stars (1) increases the 
relative strengths of emission lines, (2) decreases the 
S/N ratio, and (3) decreases the depths of many absorption features. However, 
the impact on the depths of absorption features may not be large, and the 
NaI$2.21\mu$m index changes by only a few tenths of an \AA.}
\end{figure}

	The removal of bright stars alters the co-added spectrum 
in expected ways. The amplitudes of emission lines 
are boosted when light from bright stars is removed, as the 
subtraction of star light raises the fractional contribution made by nebular 
emission to the integrated light. Noise residuals that 
result from the subtraction of bright telluric emission 
lines are also amplified, and this is most obvious at the shortest wavelengths. 
The noise in the $J-$band spectrum reflects the inherent limitations to the removal 
of telluric lines from spectra of faint objects. Nevertheless, Pa$\beta$ emission 
at $1.282\mu$m is clearly seen. Some deep absorption features 
at these wavelengths coincide with DIB transitions, although the comparisons 
in Figure 2 suggest that this may be fortuitous.

	The dominant absorption features in the $H-$band are the second-overtone 
CO bands, and the depths of these do not change substantially when the 
light from bright stars is removed. This is perhaps not surprising, as 
these features are not sensitive to stellar color (e.g. Terndrup et al. 
1991) and there is only a modest sensitivity to 
metallicity and effective temperature (e.g. Davidge 2020). 
In contrast, the depths of atomic features in the $H-$band weaken 
with the removal of light from bright stars, as expected if these lines are 
deeper in the spectra of evolved red stars.

	Line indices measure the strengths of features in a quantifiable 
manner. Indices that measure the equivalent widths of NaI$2.21\mu$m, 
CaI$2.26\mu$m, and CO(2,0) in Region 2 using the continuum and line wavelength 
intervals defined by Frogel et al. (2001) are listed in Table 2. Two entries are 
shown for each index: those measured from the spectra with only light from 
foreground stars subtracted, and those measured after light from all bright sources 
is subtracted, with the latter measurements shown in brackets. The broad wavelength 
coverage of these indices makes them susceptible to contamination from atomic and 
molecular transitions that differ from those they are intended to explore. 
For example, Carr et al. (2000) show that the NaI$2.21\mu$m index is 
contaminated by Sc, V, and CN transitions. While there is then potential ambiguity 
in the features that are monitored, these indices still measure the depths of 
features of astrophysical interest in a uniform manner.

\begin{table*}
\begin{center}
\begin{tabular}{cccc}
\tableline\tableline
Region & NaI$2.21\mu$m\tablenotemark{a} \tablenotemark{b} & CaI$2.26\mu$m\tablenotemark{a} \tablenotemark{b} & CO(2,0)\tablenotemark{a} \tablenotemark{c} \\
 & (\AA) & (\AA) & (\AA) \\ 
\tableline
1 & 2.32 (2.05) & 3.26 (3.23) & 16.55 (14.19) \\
2 & 3.04 (2.69) & 3.12 (2.07) & 16.55 (11.11) \\
3 & 3.28 (3.12) & 3.49 (3.07) & 15.10 (13.89) \\
4 & 2.38 (2.09) & 2.91 (2.50) & 12.65 (11.84) \\
\tableline
\end{tabular}
\end{center}
\tablenotetext{a}{All indices are in the instrumental system. 
Measurements made after 
light from bright stars is removed are shown in brackets.}
\tablenotetext{b}{Estimated $2\sigma$ uncertainty due to random errors is $\pm 0.03$ 
\AA.}
\tablenotetext{c}{Estimated $2\sigma$ uncertainty due to random errors is $\pm 0.15$ 
\AA.}
\caption{Na, Ca, and CO Indices in Regions 1 -- 4.}
\end{table*}

	The indices in Table 2 are in an instrumental system, and so can not be 
compared directly with the measurements made by Frogel et al. (2001). The 
F2 spectra have a lower wavelength resolution than the spectra used by Frogel 
et al. (2001), and so it is likely that the indices measured from the F2 spectra are 
systematically smaller than those in the Frogel et al. (2001) system. 
The estimated $2\sigma$ random uncertainties in these 
indices were estimated from column-to-column variances in the unbinned spectra 
coupled with the dispersion in signal within the continuum pass 
bands, and these are $\pm 0.03$\AA\ for NaI and CaI, and $\pm 0.15$\AA\ for 
CO(2,0). Given these uncertainies, there is thus a 
clear tendency for the equivalent widths measured from the 
spectra after the subtraction of bright stars to be systematically smaller than if 
only light from foreground stars is removed. Of the three indices 
considered, the NaI$2.21\mu$m index is the most robust in 
terms of sensitivity to bright stars, changing by only a few tenths of an \AA\ when 
the light from bright stars is subtracted. This conclusion also 
holds for NaI$2.21\mu$m indices measured in the other regions (see below).
 
\subsection{Comparing SEDs Along the Slit}

	The $H$ and $K$ spectra of the four regions are compared in Figures 4 
(light from bright stars included) and 5 (light from bright stars removed). We 
caution that the spectra in Figure 5 are likely not representative of composite 
stellar systems, as light from the brightest members has been removed, with the 
result that not all evolutionary states are sampled. The spectra of Regions 1 and 4 
are noisier than those of Regions 2 and 3 due to lower light levels outside of the 
central regions of the NSC. The $J$ band spectra of all regions except for Region 2, 
which is shown in Figure 3, have a low S/N ratio because of the high extinction 
towards the NSC, and so spectra at these wavelengths are not shown. 

	While these data were recorded at wavelengths where the extinction is 
greatly reduced when compared with shorter wavelengths, angular variations 
in extinction will still affect the light that is sampled. In particular, 
light from areas that have very high levels of foreground extinction will originate 
mainly from the foreground disk, which has a stellar content that likely differs 
from that of the inner Galaxy. As the foreground disk has a lower surface brightness 
than the central regions of the galaxy then these areas have conspicuous 
low light levels. Figure 1 of Davidge (1998) and Figure 32 of Nogueras-Lara 
et al. (2018b) demonstrate the variable nature of extinction near the 
NSC, and there are obvious areas where light originating from the inner 
regions of the Galaxy is blocked. The large-scale angular variations in line of 
sight extinction that are seen in the images presented by Davidge (1998) and 
Nogueras-Lara et al. (2018b) are not expected to be an issue 
for the present study, as the slit was positioned to avoid these. 
The areas of highest obscuration tend to occur to the south of the GP, which is 
where the background slit position is located. In contrast, the area sampled by the 
two F2 slit positions avoids large-scale areas of high extinction. While 
parts of Regions 3 and 4 come close to pockets of high extinction, obscuration is 
not evident in the light profile along the slit, confirming that areas of 
obscuration have been avoided.

	Extinction maps constructed by Nogueras-Lara et al. 
(2018b) from the photometric properties of red stars indicate that 
there are extinction variations over angular scales $\leq 10$ arcsec. Avoiding 
these small pockets of extinction is problematic although if -- as indicated by the 
extinction maps of Nogueras-Lara et al. (2018b) -- they occur in all four regions 
then they will bias the light along the entire slit by similar amounts. 
In fact, areas with very low light levels are found in isolated areas along the slit,
and these have small angular extents, in agreement with 
the Nogueras-Lara et al. (2018b) extinction maps. The presence of such 
pockets of high extinction serves to diminish the total 
area covered in each region, rather than bias the total light coverage. 
Moreover, the affect of such regions in the current study
is mitigated to some extent by the use of two F2 slit positions.

	All four regions sample substantial amounts of light that originates 
from NSC stars. The light profile shown in Figure 14 of Sch\"{o}del et al. (2014) 
indicates that Regions 1 and 4 each sample $5 \times 10^4$ L$_{\odot}$, while Regions 
2 and 3 sample $1 - 2 \times 10^5$ L$_{\odot}$. These estimates are comparable to the 
light originating from a globular cluster. Furthermore, while there 
is contamination from non-NSC stars, the dense nature of the NSC means that 
contamination from these objects is modest in all four regions. This is 
evident from Figures 8 and 10 of Sch\"{o}del et al. (2014) which demonstrate 
that the NSC in the areas sampled by the F2 slit is at least $5 \times$ 
brighter than non-NSC stars. Hence, while Regions 1 and 4 are more susceptible to 
contamination from non-NSC stars than Regions 2 and 3, the light in these regions 
is still dominated by NSC stars.

	Region-to-region differences in the spectra are most noticeable in the 
K-band. The differences in the emission lines reflect the complex 
distribution of ionising sources and obscuring material throughout the inner 
Galaxy (e.g. Sch\"{o}del et al. 2018; Ciurlo et al. 2016; Paumard et al. 2004; Ekers 
et al. 1983). Region-to-region variations in the $K-$band absorption spectrum 
have been quantified using the NaI$2.21\mu$m, CaI$2.26\mu$m, and 
CO(2,0) indices, and these measurements are shown in Table 2. 
Systematic behaviour is seen in the NaI$2.21\mu$m index in the 
sense that the index is larger in Regions 2 and 3 than in Regions 1 and 4. 
This trend is also seen when the spectra of the two slit positions are 
considered separately.

	A comparison of the equivalent widths in Table 2 with the amplitudes of 
emission lines in Figure 4 indicates that the depths of absorption 
features do not vary in concert with emission 
line strength. This suggests that nebular continuum emission 
does not veil absorption features in the integrated spectrum. The absorption 
spectrum is then a viable probe of the stellar content.

\begin{figure}
\figurenum{4}
\epsscale{1.0}
\plotone{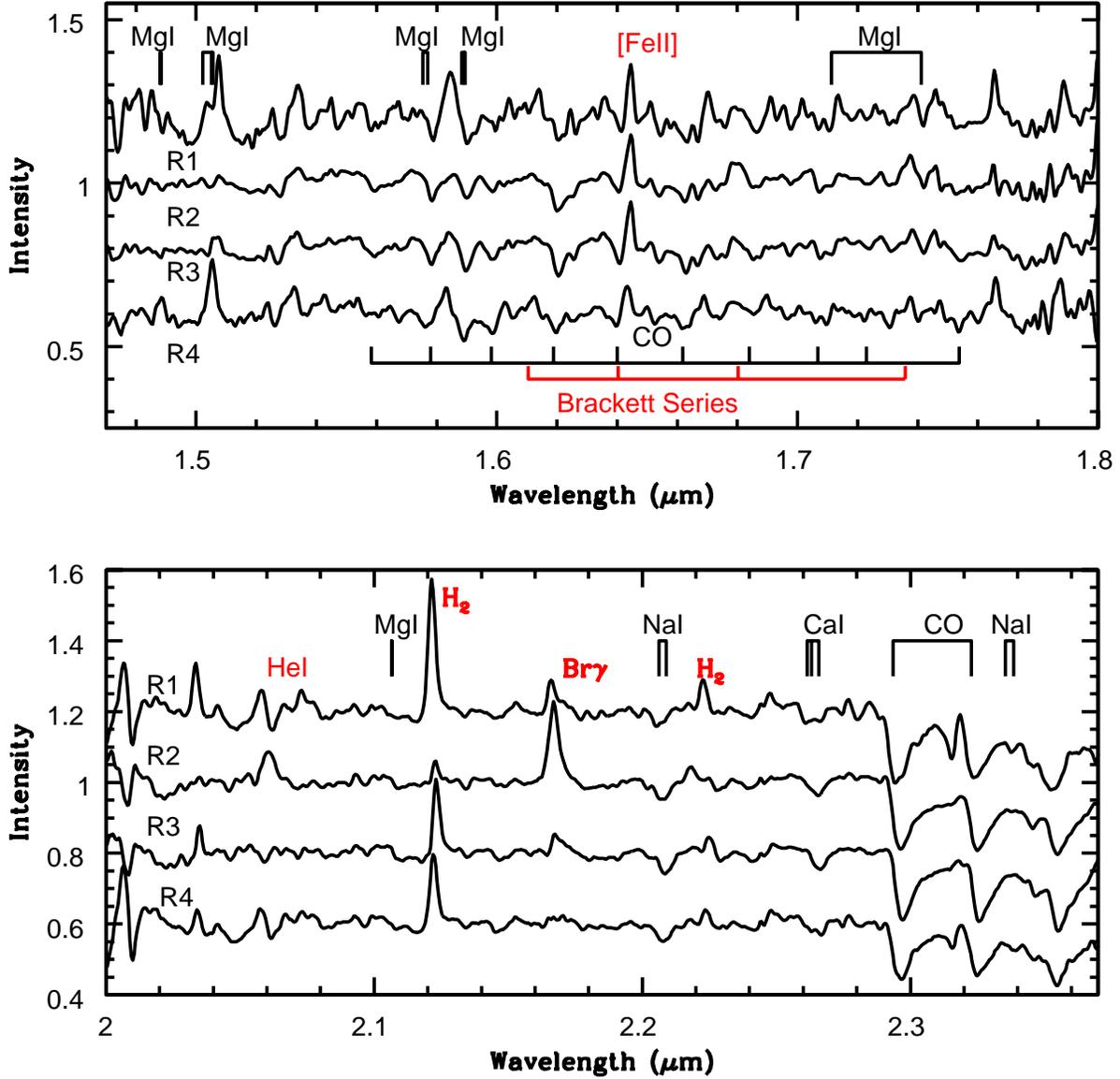}
\caption{$H$ and $K$ Spectra of the four regions with light from bright resolved 
stars included. The spectra have been offset vertically for display purposes. 
Low S/N ratios prevent obtaining useable spectra in the $J-$band for three of the 
four regions, and the $J$ spectrum of Region 2 is shown in Figure 3. All four 
regions sample part of the NSC, while Regions 2 and 3 also sample the GC Mini-spiral. 
The equivalent widths of emission lines vary from region-to-region, 
reflecting the complex distribution of ionizing sources in the NSC. Emission line 
strength and the depth of absorption features appears 
not to be correlated, suggesting that nebular continuum 
emission does not contribute significantly to the light at these wavelengths -- 
the absorption spectrum thus probes the characteristics of the stellar 
content. The deepest absorption features tend to occur in Regions 2 and 3.}
\end{figure}

	If recent star formation near SgrA* has 
proceeded for longer than a few Myr (e.g. Habibi et al. 2017) then 
some of the NIR light from Regions 2 and 3 will originate 
from a population of red objects that are cooler and have a lower surface gravity 
than the stars in Regions 1 and 4, with IRS 7 the most extreme example of such an 
object. Such evolved supergiants tend to have deep 
NIR absorption features. In fact, the equivalent width of NaI$2.21\mu$m 
is largest in Regions 2 and 3. However, the CaI$2.26\mu$m and CO(2,0) indices 
do not show a radial dependence with location along the F2 slit, suggesting 
that RSGs are not the dominant contributor to the light, which is consistent with 
the luminosity-weighted age estimate deduced from the SFH in Section 4.1. 

	Figure 5 shows spectra in which light from bright stars has been removed, 
and indices measured from those spectra are the entries in brackets 
in Table 2. There is a systematic variation in the depth of 
absorption features in Figure 5 and Table 2, in the sense that NaI$2.21\mu$m 
has a larger equivalent width in Regions 2 and 3 than in Regions 1 and 4, 
paralleling what was found in the spectra that included light from bright stars. 
That NaI$2.21\mu$m is deeper in the central regions of the NSC is thus a 
robust result. In contrast, Region 1 has the largest CaI$2.26\mu$m and 
CO(2,0) indices when light from bright stars is removed. 

\begin{figure}
\figurenum{5}
\epsscale{1.0}
\plotone{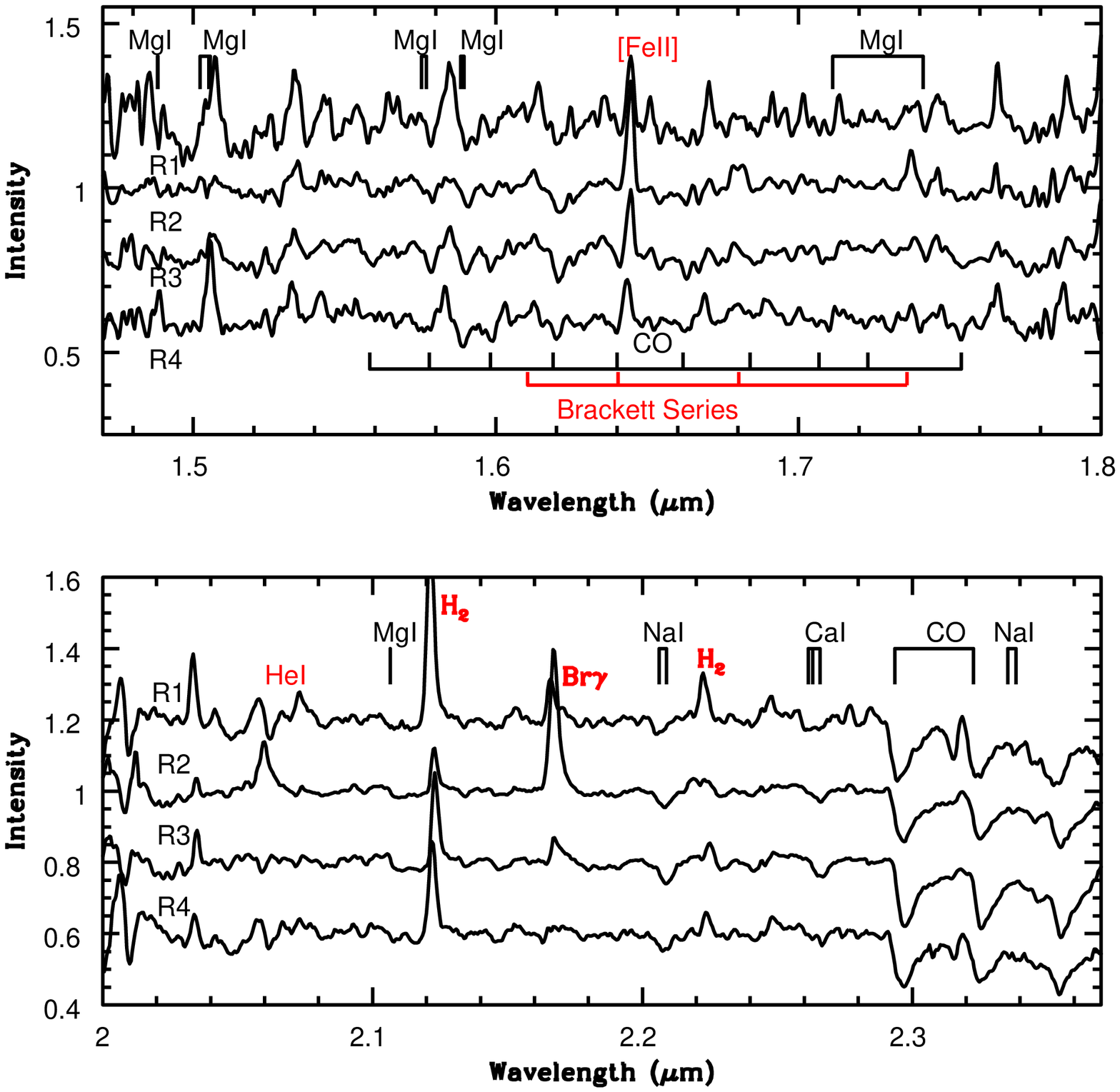}
\caption{Same as Figure 4, but showing spectra in which light from the brightest 
stars has been removed.}
\end{figure}

\section{COMPARISONS WITH MODEL SPECTRA}

\subsection{The Models}

	There has been much recent work on the development of model 
spectra of simple stellar populations (SSPs) in the NIR (e.g. 
Maraston 2005; Conroy \& van Dokkum 2012; Rock et al. 2016). However, there remain 
uncertainties in the physics of the highly evolved stars that contribute a 
significant fraction of the NIR light, and these uncertainties can have a profound 
impact on the SFH deduced solely from NIR spectra (e.g. Dahmer-Hahn et al. 2018). 
The inclusion of evolution in interacting binary systems can also significantly 
affect age estimates (e.g. Stanway \& Eldridge 2018).

	The library of available stellar spectra is of 
obvious importance when generating model spectra. The IRTF 
library compiled by Rayner et al. (2009) is the basis 
for many SSP models, and in its original form it did 
not include stars with spectral-types earlier than A0 or a large set of objects with 
non-solar metallicities and chemical mixtures. These limitations complicated 
efforts to model the NIR SEDs of systems that are younger than a few tenths of a Gyr 
and/or that have experienced chemical enrichment histories that differ from the solar 
neighborhood. Having access to spectra that cover a full range of the highly evolved 
objects that can contribute significantly to the NIR light is also important. 
For example, the spectra of C stars in the IRTF library may not track the full 
suite of C star characteristics (Davidge 2020). The IRTF library is being expanded 
to cover a larger parameter space (Villaume et al. 2017), while other independent 
libraries are also becoming available (e.g. Gonneau et al. 2020).

	The models used here are from the E-MILES compilation (Rock et al. 2016), and 
were generated from the BaSTI isochrones (Pietrinferni et al. 2004; Cordier et 
al. 2007) with spectra at NIR wavelengths constructed from stars in the 
IRTF library. A Chabrier (2001) mass function has been adopted. Paumard et al. 
(2006), Bartko et al. (2010) and Lu et al. (2013) conclude that the mass function of 
young massive stars near the GC is flatter than that among similar stars in 
the Solar neighborhood. However, the luminosity-weighted age towards the center 
of the NSC is a few Gyr (see below). The mass function of recently formed stars 
is then not expected to be a major issue for the current study, as most of the light 
comes from stars with a lower mass than those examined by Paumard et al. (2006), 
Bartko et al. (2010) and Lu et al. (2013). The stars that dominate the light from 
systems with ages of a few Gyr have also likely been well-mixed throughout the 
NSC -- any localized variations in the mass function would then have been 
smoothed out.

	A forward modelling approach is adopted for 
this study. Luminosity-weighted ages and metallicities 
have been found using results from studies of resolved stars near the GC. 
Figure 14 of Pfuhl et al. (2011) compares SFR measurements for the 
GC from various studies, and a hand-fit SFR $vs$ time relation from the entries 
in that figure, which is weighted towards old and intermediate ages, 
was adopted for computing luminosity-weighted ages.

	While there have been other age estimates made since the Pfuhl et 
al. (2011) compilation was published, these are not expected to alter our results. 
For example, Lu et al. (2013) estimate an age of 2.8 or 3.9 Myr for the 
most recent episode of star formation, whereas the youngest entry in the Pfuhl et 
al. (2011) compilation is $\sim 6 - 7$ Myr. However, given that the 
weighting of old and intermediate populations in the NIR is greater than that of 
young populations then the luminosity-weighted age is not sensitive to changes 
in age estimates for the youngest stars.

	There is some uncertainty in the ages of bulge stars. 
Renzini et al. (2018) discuss recent photometric studies of bulge stars, and note 
that these tend to find a uniformly old age. In contrast, age estimates based 
on spectroscopic properties yield a significant intermediage age component among 
metal-rich bulge stars (e.g. Bensby et al. 2017). While this introduces 
uncertainty in the luminosity-weighted age, this should not 
affect the task at hand, as the NIR SEDs of systems 
older than a few Gyr change only slowly with age. In the next section it is 
shown that the NaI$2.21\mu$m and CaI$2.26\mu$m indices 
have only a modest sensitivity to system age.

	The integrated NIR luminosity of stars in various time 
intervals was found by assigning a M/L measurement to different age 
intervals. Figure 24 of Maraston (2005) summarizes $K-$band M/L ratios 
from various studies. There are not substantial model-to-model variations in the M/L 
ratios in that figure except near $\sim 1$ Gyr. This age is where the fractional 
contribution from AGB stars to the NIR light is expected to be significant, and 
the dispersion in M/L ratios at this age in Figure 24 of Maraston (2005) are 
due to different approaches that were adopted to track AGB evolution. The 
uncertainties in M/L ratios near 1 Gyr do not introduce major uncertainties when 
generating luminosity-weighted ages in the NIR as the SFR 
in Figure 14 of Pfuhl et al. (2011) is comparatively low near 1 Gyr. 
Adopting the Pfuhl et al. (2011) SFH and Maraston (2005) M/L ratios then 
the luminosity-weighted age along the SgrA sight line is $\sim 3$ Gyr. 

	As for luminosity-weighted metallicity, there is evidence 
for a preponderance of super-solar metallicity stars near the center of the 
NSC, although with a broad range of metallicities. Rich et al. (2017) 
discuss metallicity estimates of NSC giants and find 
a moderate spread in metallicity among older stars, with a mean [Fe/H] 
$\sim -0.3$. Cunha et al. (2007) find that young stars near the GC have a mean 
[Fe/H] $\sim +0.1$. Feldmeier-Krause et al. (2017b) examine the spectra of 
stars in the central 4 pc$^2$, and find a wide distribution with a peak near 
[M/H] $= +0.3$ dex, while Nandakumar et al. (2018) also find a peak metallicity 
$+0.3$ dex among stars near the GC. The stars observed for the latter study have $K$ 
between 11 and 12, and are thus among the objects that are subtracted from 
the spectra throughout this work to examine luminosity effects. Schultheis 
et al. (2019) compile results from a number of studies and conclude 
that stars near the GC have super-solar metallicities while those in the inner 
bulge have a peak metallicity near solar. Given this diversity, models are 
considered here that have slightly super-solar metallicities, although slightly 
sub-solar metallicity models are also considered to examine the dependence 
of the results on the assumed metallicity, as well as possible systematic effects 
in metallicity estimates that may occur among M giants (Thorsbro et al. 2018). 
A scaled-solar chemical mixture is assumed. 

\subsection{Comparisons with Observations}

	Model spectra of a 3 Gyr SSP with [Fe/H] $= -0.26$ and $+0.06$ are compared 
with part of the $K-$band spectrum of SgrA in Figure 6. The models have been 
smoothed and re-sampled to simulate the wavelength resolution and cadence 
of the F2 spectra. The $2.2 - 2.35\mu$m wavelength interval was selected for 
comparison with the models as it contains a number of deep absorption features, 
is free of strong emission lines, 
and has a high S/N ratio when compared with other wavelengths. 

\begin{figure}
\figurenum{6}
\epsscale{1.0}
\plotone{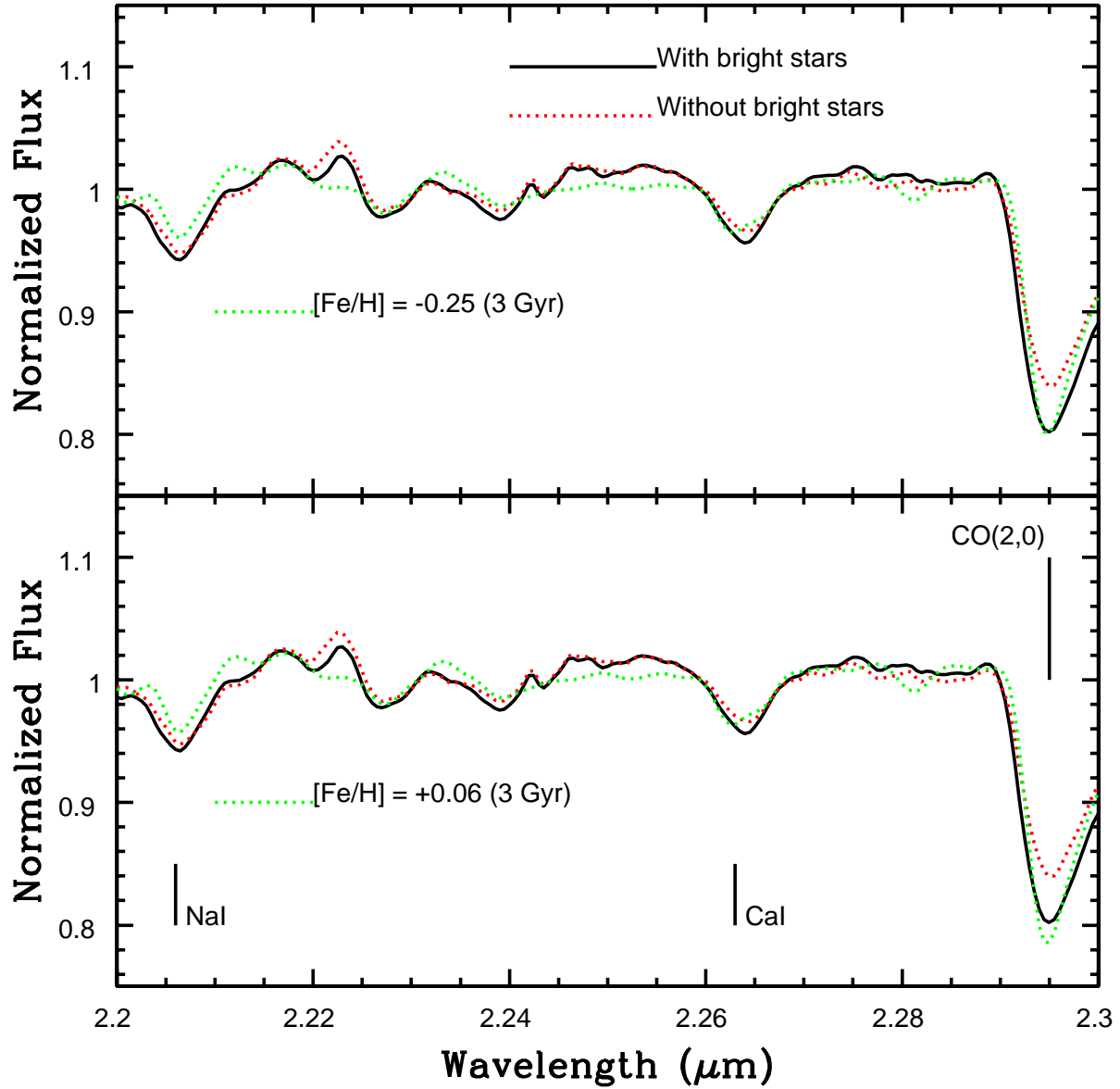}
\caption{Model spectra of SSPs with [F/H] = --0.25 (top panel) and +0.06 
(bottom panel) and an age of 3 Gyr are compared with the mean spectrum of Regions 2 
and 3. The black line shows spectra in which light from bright stars has 
been retained, while the red dashed line shows the spectrum with this light 
removed. The $2.2 - 2.35\mu$m wavelength interval is free of strong emission 
lines, and so is well-suited for comparisons with model spectra.}
\end{figure}

	NaI$2.21\mu$m, CaI$2.26\mu$m, and CO(2,0) indices 
were measured from the model spectra, and the results 
are listed in Table 3. Indices were measured from 1, 3, and 10 Gyr models wih 
[Fe/H] =--0.25 and +0.06 to allow age and metallicity effects to be assessed. The 
entries in Table 3 indicate that the NaI$2.21\mu$m and CaI$2.26\mu$m indices are 
only mildly sensitive to changes in metallicity and age at the wavelength resolution 
of the F2 spectra, producing differences of only a few tenths of an \AA\ 
over the age and metallicity ranges considered. Any differences between the 
observations and models larger than a few tenths of an \AA\ are then likely not due 
to uncertainties in age and/or overall metallicity. The CO(2,0) index is more 
sensitive to changes in age and metallicity than either NaI$2.21\mu$m and 
CaI$2.26\mu$m. Indices were also measured in spectra with 
$\frac{\lambda}{\Delta \lambda} \sim 400$, which is lower than the resolution of 
650 in the F2 $K-$band spectra, and the NaI$2.21\mu$m and CaI$2.26\mu$m indices 
measured from the lower resolution spectra differ by only a few 
hundredths of an \AA\ from those in Table 3.

\begin{table*}
\begin{center}
\begin{tabular}{cccc}
\tableline\tableline
Model & NaI$2.21\mu$m\tablenotemark{a} \tablenotemark{b} & CaI$2.26\mu$m\tablenotemark{a} \tablenotemark{b} & CO(2,0)\tablenotemark{a} \tablenotemark{c} \\
 & (\AA) & (\AA) & (\AA) \\ 
\tableline
--0.25 1 Gyr  & 2.08 & 2.01 & 11.58 \\
+0.06 1 Gyr & 2.13 & 2.12 & 12.04 \\
 & & & \\
--0.25 3 Gyr  & 2.16 & 2.18 & 13.27 \\
+0.06 3 Gyr & 2.31 & 2.35 & 14.18 \\
 & & & \\
--0.25 10 Gyr  & 2.29 & 2.40 & 14.05 \\
+0.06 10 Gyr & 2.44 & 2.46 & 14.25 \\
 & & & \\
NGC 4491 & 2.62 & 2.18 & 15.58 \\
\tableline
\end{tabular}
\end{center}
\tablenotetext{a}{All indices are in the instrumental system. Those made with 
light from bright stars removed are shown in brackets.}
\tablenotetext{b}{Estimated $2\sigma$ uncertainty due to random errors is $\pm 0.03$ 
\AA.}
\tablenotetext{c}{Estimated $2\sigma$ uncertainty due to random errors is $\pm 0.15$ 
\AA.}
\caption{Na, Ca, and CO Indices in Model Spectra and NGC 4491}
\end{table*}

	Previous efforts to model the NaI$2.21\mu$m and CaI$2.26\mu$m features in 
the spectra of massive spheroids underestimated their strengths (e.g. Alton et al. 
2018). Galaxy-to-galaxy variations in the equivalent width of NaI$2.21\mu$m 
have also been found (e.g. Rock et al. 2017, and others), 
suggesting that this index may be influenced by parameters other 
than age and metallicity. Although deep NaI$2.21\mu$m features 
are most commonly associated with massive spheroidal galaxies, these are also 
seen in lower mass early-type disk galaxies (Davidge 2020). 
Differences in [Na/Fe], the nature of the mass function at the lower 
end of the main sequence, and variations in [C/Fe] (e.g. Rock et al. 2017; 
Sarzi et al. 2018) have been forwarded as possible causes of the galaxy-to-galaxy 
variations in the NaI$2.21\mu$m feature in integrated spectra. 

	The NaI$2.21\mu$m equivalent widths obtained from spectra of 
Regions 1 and 4 in which light from bright stars is included 
fall within the range defined by the models. 
However, the NaI$2.21\mu$m indices in Regions 2 and 3 are 
0.7 -- 1.0\AA\ larger than in the 3 Gyr models. The CaI$2.26\mu$m 
indices in all four regions are uniformly larger than in the 
models, which is consistent with what is seen in other galaxies 
(e.g. Alton et al. 2018; Davidge 2020). When light from bright stars is removed 
from the spectra then the NaI$2.21\mu$m indices for Regions 2 and 3 are still 0.4 -- 
0.8\AA\ deeper than in the 3 Gyr models. The equivalent width of 
CaI$2.26\mu$m also drops when the light from bright stars is removed, but is still 
larger than predicted by the models for three of the four regions. In summary, the 
differences between the depths of the CaI$2.26\mu$m and NaI$2.21\mu$m features 
and the model measurements can not be attributed solely to stars with $K < 12$. 

	With the exception of Region 4, the CO(2,0) indices measured 
from the F2 spectra with light from bright stars included are uniformly deeper 
than those measured from the SSP model spectra. 
When light from bright stars is removed then the CO(2,0) indices are 
consistent with those in the model spectra. The bright stars that were 
removed when constructing the SgrA spectrum in Figure 6 are the most luminous, 
highly evolved objects in the field, and thus typically should have deep CO bands, 
although we recall that the CO indices in IRS 7 are {\it weaker} than found in 
$\alpha$ Ori. To the extent that the spectra can be 
represented by a luminosity-weighted age and metallicity 
then the SSP models considered here thus underestimate 
the contribution from CO(2,0) in the spectrum of bright stars. An alternative is 
to assume that the spectra have a luminosity-weighted metallicity that is 
$\sim 0.6 - 0.7$ dex, but this is not consistent with the metallicity 
estimates for individual stars discussed in Section 3. A possible explanation for 
these results is that the chemical mixture of stars in the NSC may not match 
those in the solar neighborhood.

	Sellgren et al. (1987) and Blum et al. (1996) find that the NaI$2.21\mu$m and 
CaI$2.26\mu$m absorption features are deeper in 
the low resolution spectra of stars in the NSC than in the 
spectra of solar neighborhood counterparts, and the measurements made from the 
F2 spectra are consistent with this. While this might point to super-solar 
[Na/Fe] and [Ca/Fe] in these stars, Carr et al. (2000) find 
that the NaI$2.21\mu$m index is contaminated by 
lines of V, Sc, and CN at low wavelength resolutions, and that it is those lines that 
are driving differences in the depths of the NaI$2.21\mu$m index in IRS 7 and 
$\alpha$ Ori. This opens the possibility that 
elements other than Na might explain the difficulty matching the depths of 
NaI$2.21\mu$m (Alton et al. 2018, but see also Rock et al. 2017) in the spectra of 
early-type galaxies.

	The comparisons in Tables 2 and 3 indicate that 
deep NaI$2.21\mu$m absorption is not restricted to cool giants in the NSC, but is 
also seen in integrated light that samples a large range of 
evolutionary states and effective temperatures, 
as the NaI$2.21\mu$m and CaI$2.26\mu$m indices are larger than expected even after 
light from luminous bright stars is removed. NaI$2.21\mu$m 
and CaI$2.26\mu$m equivalent widths that are larger than predicted by the 
models must then occur in stars that span a wide range of evolutionary 
states. This could occur if there are systematic differences in chemical mixture 
between stars near the center of the NSC and in the solar neighborhood. 

	The existence of non-solar chemical mixtures among GC stars is a matter 
of controversy. Do et al. (2018) find deeper than expected Sc lines in the 
spectra of M giants in the NSC, and interpret this as evidence for 
a non-solar chemical mixture. However, Thorsbro et al. (2018) argue that the 
formation of these lines in the photospheres of cool stars
is not well understood. Nevertheless, the differential comparison 
conducted by Carr et al. (2000) of the spectra of IRS 7 and $\alpha$ Ori, which 
are stars that have similar effective temperatures and surface gravities, 
point to clear differences in the depths of lines of Sc, V, and CN.

	Non-solar chemical mixtures would indicate that 
the bulk of the stars along this line of sight formed from 
material that did not experience evolution in a thin disk-like 
environment. Rather, they may have formed from material that was enriched very 
quickly, with only minimal subsequent dilution over long time scales. 
Alternatively, it could also indicate an IMF that differs from that 
in the solar neighborhood. 
The Na abundance of stars in globular clusters show systematic effects that 
differ from those of stars in open clusters (e.g. Gratton et al. 2012; Cunha et 
al. 2015), and the chemical evolution of globular clusters might then yield 
clues for understanding the non-solar mixtures that may occur near the GC.

\section{THE SPECTRUM OF THE CENTER OF THE NSC}

\subsection{Comparisons with the Spectra of Other Galaxies}

	Neumayer et al. (2020) review the properties of the Galactic NSC in 
the context of nuclear clusters in other galaxies.
They conclude that the NSC is a typical nuclear cluster 
in terms of spatial extent, mass, and shape. The stellar 
content of the NSC is also consistent with that of other 
late-type galaxies, in the sense that there is a broad range of ages that 
extend from young to old populations.

	The spectrum of the NSC contains a fossil record of its SFH. Given 
that the NSC has similar properties to the nuclear clusters in other galaxies 
then it is perhaps not surprising that the integrated NIR spectrum of the NSC has 
characteristics that are similar to those of nuclei in other galaxies. 
Evidence to support this claim can be found in Figures 7 and 8 where the mean 
spectrum of Regions 2 and 3 (hereafter referred to as the SgrA spectrum) 
is compared with the nuclear spectra of the Sc galaxy NGC 7793 (Davidge 2016), 
and the star burst galaxy NGC 253 (Davidge 2016). These galaxies represent 
two distinct types of late-type galaxies. NGC 7793 
is a nearby Sc galaxy, with on-going star formation throughout its disk. 
While the nucleus of NGC 7793 contains stars younger than $\sim 100$ Myr 
(Kacharov et al. 2018), its NIR spectrum lacks prominent emission lines. 
The presence of an inverse color gradient also hints at a 
a complicated formation history for its nucleus 
(Kacharov et al. 2018). NGC 253 is a barred Sc galaxy, and is one of the nearest 
starburst galaxies, with vigorous star formation occuring throughout. 

	Nuclear star clusters are found over a broad range of Hubble 
types (e.g. Seth et al. 2008). To enable a comparison with a nuclear 
cluster that is not in a late-type disk galaxy, spectra of the Virgo cluster 
early-type disk galaxy NGC 4491 (Davidge 2020) is also shown in Figures 
7 and 8. NGC 4491 is an SBa(s) galaxy in the Virgo cluster that has little or 
no on-going star formation in its disk, but has a blue nuclear SED 
at visible wavelengths and a red nuclear SED in the MIR (Davidge 2018). 
It is the only early-type galaxy in the sample discussed by Davidge (2020) 
that has NIR line emission, and has a near-solar central metallicity 
(Davidge 2018). Moreover, the NaI$2.21\mu$ and CaI$2.26\mu$m features 
in the NGC 4491 spectrum are deeper than predicted by models, as is the case 
in Regions 2 and 3.

	The spectra in Figures 7 and 8 sample a range of 
spatial scales that exceed the area covered with the F2 spectra. 
The NGC 253 spectrum covers a region with an 8 parsec radius, while the NGC 7793 
spectrum samples an area with a 16 parsec radius. The NGC 4491 spectrum 
covers an $80 \times 80$ parsec area.

	The $H-$band spectra in Figure 7 have a range of characteristics. The 
emission spectrum of NGC 253 IRC in the $H-$band is dominated by [FeII] and Brackett 
lines, and these have much larger equivalent widths than in the other nuclei, 
indicating a much younger luminosity-weighted age. 
Still, the absorption features in the NGC 253 IRC $H-$band spectrum 
match those in the other systems. There is weak line emission in the NGC 7793 
$H-$band spectrum, indicating that there has not been extensive levels 
of star formation within the past few Myr in the 
central few parsecs of that galaxy. The depths of absorption features 
in the H$-$band spectrum of NGC 7793 match those in the SgrA spectrum.
Emission lines in the NGC 4491 spectrum are intermediate in strength between 
those in NGC 253 and NGC 7793, and come closest to matching those in 
SgrA. The $H-$band absorption spectra of NGC 4491 and SgrA spectra also agree.

\begin{figure}
\figurenum{7}
\epsscale{1.0}
\plotone{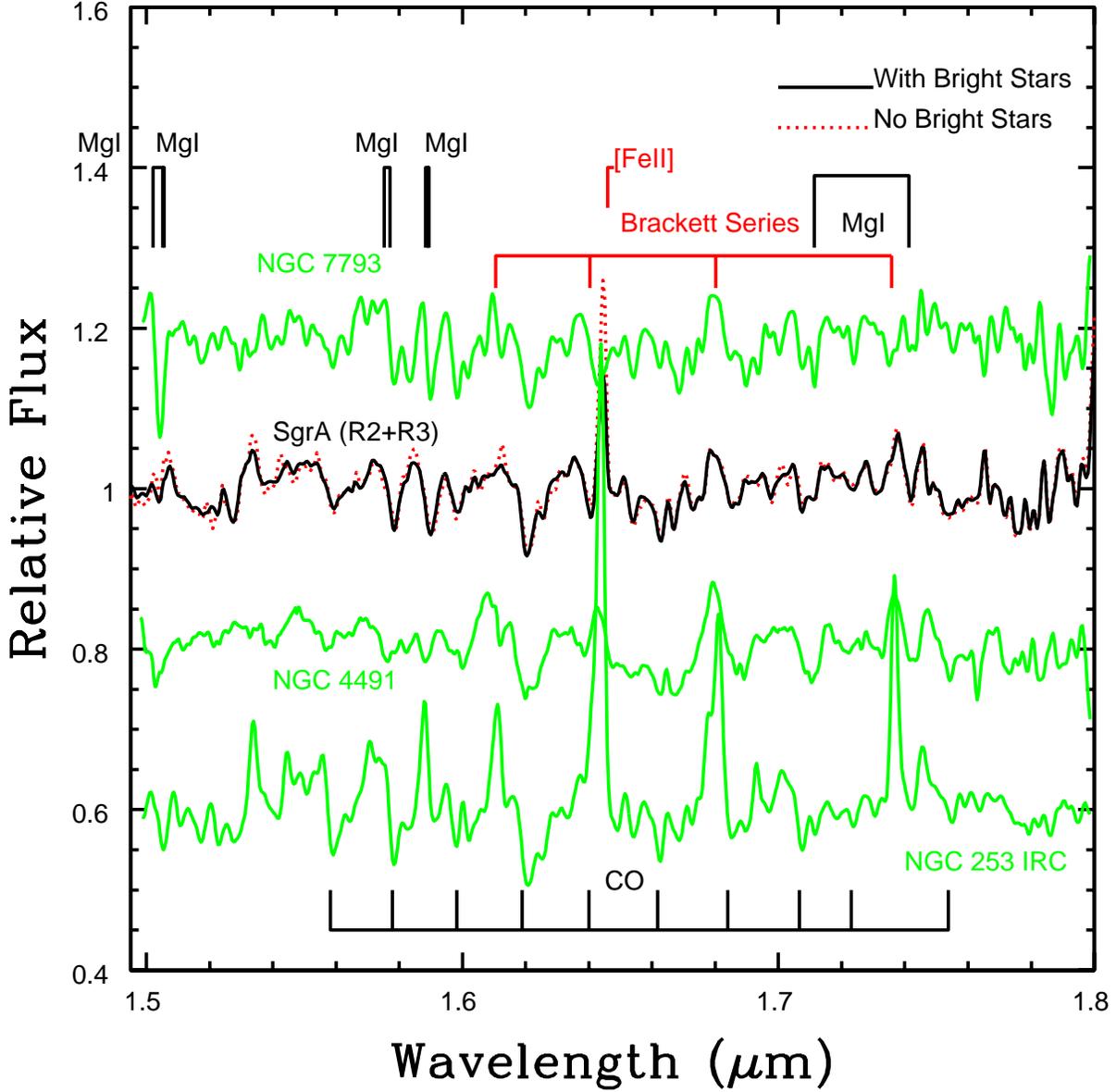}
\caption{Comparing the mean $H-$band spectrum of Regions 2 and 3 (hereafter 
refered to as the `SgrA spectrum') with the spectra 
of extragalactic nuclei that host current or recent star formation. 
The black line shows the continuum-normalized SgrA 
spectrum with light from bright stars retained, while 
the dashed red line shows the spectrum with light 
from bright stars removed. The green lines show the spectra of 
NGC 7793 (Davidge 2016), NGC 253 IRC (Davidge 2016), and NGC 4491 (Davidge 
2020). The SgrA and NGC 4491 spectra have similar appearances.}
\end{figure}

	There are clear differences between the $K-$band spectra of the nuclei 
in Figure 8. The first-overtone CO bands in the NGC 7793 spectrum are markedly deeper 
than in the SgrA spectrum, suggesting that there is a larger fractional 
contribution from cooler, low surface gravity stars to the NIR light in the central 
few parsecs of NGC 7793 when compared with the central parsec of SgrA. The 
luminosity-weighted age in the region sampled by the NGC 7793 spectrum is thus 
substantially younger than in the F2 NSC observations. While the 
depths of absorption lines in the SgrA and NGC 253 spectra more-or-less agree, 
the emission lines in the NGC 253 spectrum have larger equivalent widths. 

\begin{figure}
\figurenum{8}
\epsscale{1.0}
\plotone{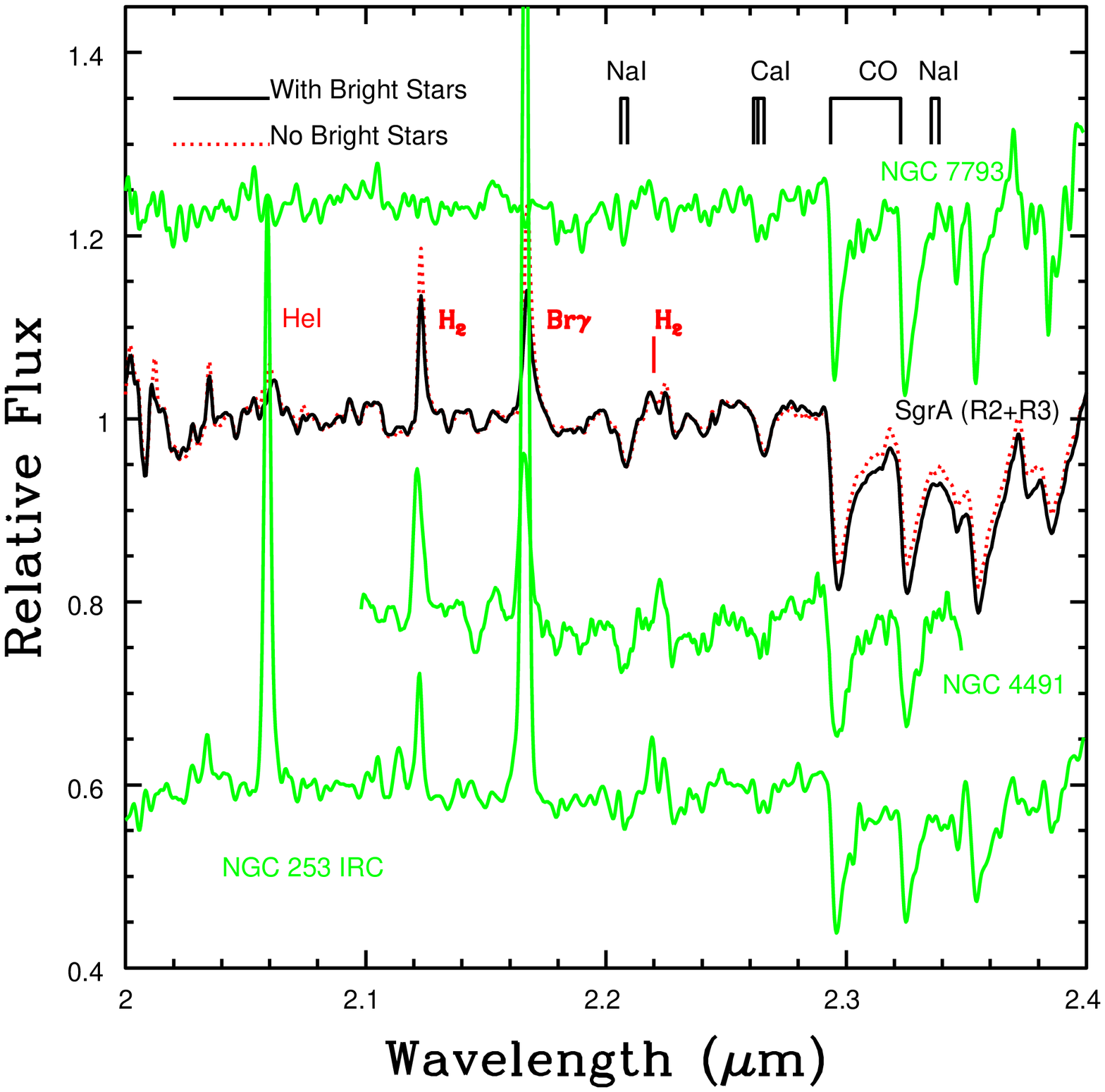}
\caption{Same as Figure 7, but with $K-$band spectra. The 
NGC 4491 spectrum spans a narrower wavelength range than the other spectra as 
it has a poor S/N ratio outside of the plotted interval. As in Figure 7, there 
is good visual agreement between the SgrA and NGC 4491 spectra.}
\end{figure}

	The similarities between the SgrA and NGC 4491 absorption spectra in Figures 
7 and 8 are not surprising considering that the luminosity-weighted age 
and metallicity estimates for the SgrA line of sight found in Section 4 
agree with those deduced from the integrated spectrum of NGC 4491 by Davidge 
(2018), who based that estimate on absorption features at visible and 
red wavelengths. The depths of the first overtone CO bands, 
NaI$2.21\mu$m and CaI$2.26\mu$m in these spectra are similar. Line indices 
measured from the NGC 4491 spectrum are listed in Table 3. The 
NGC 4491 indices tend to be smaller than those in SgrA, although the NGC 4491 
NaI$2.21\mu$m and CO(2,0) indices still exceed those measured from the models. 

	A surprising result is that the emission lines in the $K-$band 
spectra of NGC 4491 and SgrA have comparable equivalent widths. 
These similarities are noteworthy given the marked differences between 
these galaxies. SgrA is at the center of a star-forming SBbc 
galaxy that is in a more-or-less isolated environment. The Galaxy likely has not 
interacted with galaxies of similar mass, although there is 
an entourage of dwarf companions with which there likely has been interactions. In 
contrast, NGC 4491 is an intermediate mass SBa(s) disk galaxy with quenched 
star formation in the disk; Lisker et al. (2006) classify NGC 4491 as a `dwarfish 
S0/Sa'. NGC 4491 is in a crowded environment where interactions with galaxies of 
similar or larger size are almost certain to have occured. 

	Perhaps of greatest significance is that very 
different physical volumes are sampled by the SgrA and NGC 4491 
spectra; NGC 4491 is three orders of magnitude more distant than the GC, and so 
the F2 spectrum of NGC 4491 includes light from a much larger intrinsic volume, 
and hence a larger number of stars, than in the SgrA F2 observations.
Davidge (2018; 2020) discusses the star-forming environment of 
the nucleus of NGC 4491, and notes that the MIR SED of the 
central regions of NGC 4491 is similar to that of the starburst dwarf galaxy 
NGC 5253, where massive, compact clusters are forming 
(e.g. Calzetti et al. 2015, and references therein). Boselli et al. 
(2010) found that the characteristic dust temperature in its central regions is the 
highest of all galaxies in the Herschel Reference Survey. Even though 
the similarities in the NIR emission line spectra point to similar {\it mean} 
densities for the ionizing radiation field, the localized conditions in 
SgrA are likely not as extreme as those in NGC 4491.

\subsection{The Spectrum of the Young Component}

	The SFRs compiled in Figure 14 of Pfuhl et al. (2011) 
indicate that while the NIR light from the center of the NSC is dominated 
by stars with intermediate and old ages, there is a 
significant contribution from stars that formed within the past $\sim 10$ Myr. 
On a basic level, the spectra of any star-forming region can be 
thought of as the luminosity-weighted combination of light from a young stellar 
component and older background/foreground populations. An integrated spectrum of the 
young component near the GC can then be extracted by subtracting out light 
from the intermediate age and older components.

	Models with ages of 3 and 10 Gyr 
and [Fe/H] = --0.25 and +0.06 were subtracted from the 
SgrA (i.e. the mean of Region 2 and 3) spectrum to assess the affect of varying 
age and metallicity on the resulting young spectrum. Following the 
Pfuhl et al. (2011) SFR summary, these models were scaled so that the older 
population account for $\sim 70\%$ of the NIR light. 
The spectrum that results after subtracting the 3 Gyr model with [Fe/H] = --0.25
is shown in Figure 9. The spectra produced by subtracting the 10 Gyr model 
and those with [Fe/H] = 0.06 are almost identical to those shown in Figure 9.

\begin{figure}
\figurenum{9}
\epsscale{0.9}
\plotone{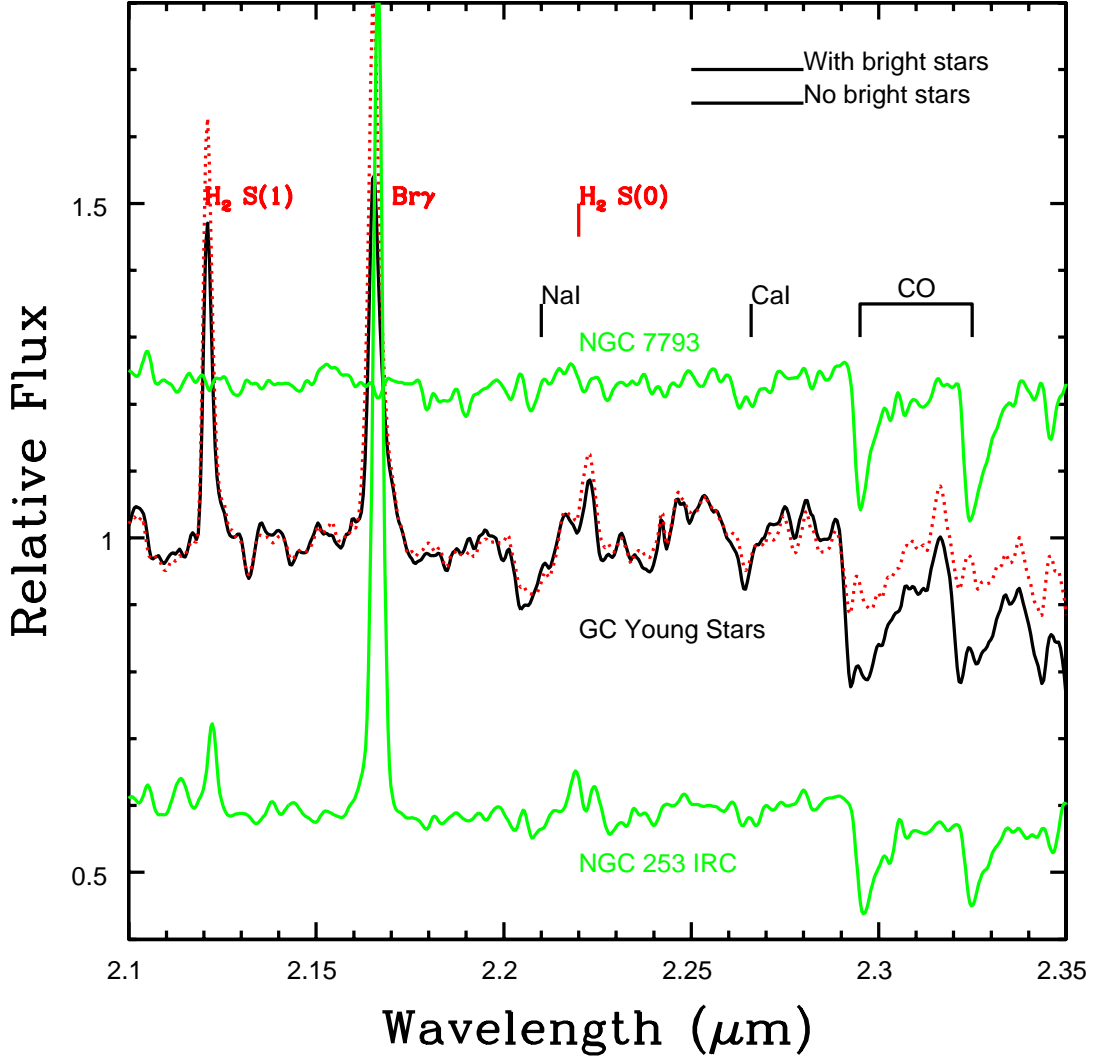}
\caption{Spectrum of the young component near the GC, which was constructed by 
subtracting a scaled SSP spectrum with [Fe/H] = -0.25 and 3 Gyr from the SgrA 
spectrum. The black line shows the differenced spectrum that results when 
light from bright stars is included, while the dashed red line 
shows the differenced spectrum when the light from bright stars is 
removed. Spectra of the center of NGC 7793 and NGC 253 from 
Davidge (2016) are also shown. The CO bands in the differenced spectrum provide 
insights into the young, cool stellar content. The deep CO bands in the 
black spectrum originate predominantly from IRS 7, although very luminous AGB stars 
that formed during the past $\sim 1$ Gyr may also contribute to these 
features. The CO bands largely disappear when light 
from bright sources is removed, indicating that the 
contribution made by cool pre-main sequence objects is modest at these wavelengths.}
\end{figure}

	The removal of light from old stars enhances the emission 
spectrum, amplifying features that are faint in the initial spectrum. 
As the spectra were recorded through slits then not all 
of the line emission is sampled. Still, the F2 slit pointings sample 
the strongest regions of Br$\gamma$ and HeI$2.06\mu$m emission shown 
in Figures 7 and 8 of Feldmeier-Krause et al. (2015). The spatial 
distribution of the Br$\gamma$ and HeI emission is also very similar. Hence, the 
SgrA emission spectrum in Figure 9 should be representative of the 
most vigorous star-forming areas near the center of the NSC.

	Br$\gamma$ and H$_2$ 1-0 S(1) have similar strengths in the 
GC spectrum in Figure 9, which is consistent with what is seen in 
star-forming galaxies (e.g. Puxley et al. 1990). 
The H$_2$ 1-0 S(1) emission line in the young SgrA spectrum has a larger equivalent 
width than in the NGC 253 spectrum. Rosenberg et al. (2013) discuss the 
source of H$_2$ emission in NGC 253. Based on the ratio of the 2-1 and 1-0 S(1) 
H$_2$ lines, coupled with the similar spatial distributions of H$_2$ and polycyclic 
aromatic hydrocarbon emission, they conclude that H$_2$ emission near the 
center of NGC 253 likely originates from B stars.

	The 2-1 S(1) line near $2.25\mu$m 
in the young GC spectrum is not clearly detected, and it is much weaker 
than the 1-0 line. As in NGC 253, the H$_2$ line strengths are thus consistent 
with H$_2$ emission powered by fluoresence, rather than 
shocks. While the comparatively large equivalent width of the 1-0 S(1) 
line in the young GC spectrum might suggest a higher luminosity-weighted density 
of lower mass ionizing stars than near the center of NGC 253, we caution 
that the F2 slit does not sample all parts of the 
star-forming region and the H$_2$ line varies in strength from region-to-region 
(e.g. Yusef-Zadeh et al. 2001).

	Deep absorption lines are seen in the young GC spectrum when 
light from bright stars is included. These deep CO bands are 
due in large part to IRS 7. Given that the SSP models that have been 
used to construct the young GC spectrum have ages well in excess of a Gyr then 
there could also be a contribution to the deep NIR absorption features in Figure 9 
from AGB stars in populations that formed a few hundred Myr in the past. Still, 
it is unlikely that much the absorption features in Figure 9 originate predominantly 
from such a population since the SFH indicates that the SFR was low 0.5 - 1.0 Gyr 
in the past. 

	Some types of pre-main sequence (PMS) stars might 
contribute to CO absorption in the integrated spectrum of a young system. 
Eckart et al. (2013) conclude that star formation near the GC is on-going, and 
Pei$\beta$ker et al. (2020) identify candidate PMS stars near the center of the NSC.
However, Nayakshin \& Sunyaev (2005) find lower than expected X-ray emission from 
the inner Galaxy, suggesting low frequency of PMS stars that could 
be consistent with a top-heavy mass function. In fact, the mass function 
exponent found by Lu et al. (2013) can explain the X-ray luminosity.

	FU Orionis objects are PMS stars that have 
a low mass and experience perodic, short-lived accretion 
events from a companion that causes a large increase in magnitude. These objects 
have deep, supergiant-like CO bands in their spectra (e.g. 
Hartmann et al. 2004; Connelley \& Reipurth 2018), and are intrinsically 
much fainter than RSGs, with M$_K \sim -1$ for the prototype star FU Orionis, 
making them more-or-less comparable to the brightness of red clump stars. 
FU Ori stars are thus within the detection 
limits of current photometric studies of resolved objects in the NSC, 
and none have been detected. Given that the outbursts 
that characterize FU Ori stars are also very rare, it is 
unlikely that they contribute significantly to the integrated NIR light.
In fact, when light from bright stars is excluded then 
the CO bands weaken, and the significance of these features 
is not high given the uncertainties in the contribution that light from 
old and intermediate age stars make to the total population. 

	While NaI$2.21\mu$m and CaI$2.26\mu$m are deep in the GC young spectrum 
after light from bright stars is removed, these are likely an artifact of the 
model spectrum having NaI and CaI features that are too weak. 
Therefore, once light from bright stars is removed then 
the young spectrum can be characterized as a mainly featureless continuum, 
as expected given the large concentration of early-type stars that 
are fainter than $K = 12$ in the inner regions of the NSC (e.g. Paumard et al. 2006; 
Do et al. 2013).

\section{DISCUSSION \& CONCLUSIONS}

	Spectra recorded with F2 on GS have been used to examine the SED of the 
NSC along its major axis in the wavelength interval $1.3\mu$m -- $2.4\mu$m. 
The 4.3 arcmin F2 slit samples stars near the GC that formed 
within the past few Myr, as well as stars in the NSC, the NB, and the 
intermediate age/old stellar substrate of the inner regions of the Galaxy. The 
slit pointings cover roughly 100 arcsec$^2$ in the central parts of the NSC, while an 
even larger area is covered in the outer regions of the NSC. 

	The study of integrated light has the potential to allow the full range of 
stellar content in a system to be examined, including 
intrinsically faint objects that can not be detected 
individually due to crowding and/or faintness, but that still contain important 
clues into the evolution of the NSC and the NB. While there are 
possible issues understanding the formation of lines in the NIR spectra of cool 
stars (e.g. Thorsbro et al. 2018), the NIR is by default the primary wavelength 
region for probing the stellar content of the NSC as spectra at visible 
wavelengths are unattainable. Studies of the integrated NIR spectrum of the NSC also 
enable direct comparisons with the central regions of other galaxies, allowing 
the nature of the inner regions of the Galaxy to be compared 
with those of other galaxies. For example, the detection of emission lines in the 
SgrA spectrum that have comparable equivalent widths to those in other galaxies 
would be consistent with similar characteristics in the local ionizing 
radiation field. 

	The conclusions are as follows:

\vspace{0.3cm}
\noindent{1)} The depths of absorption features in the integrated NIR spectrum of 
the NSC are {\it roughly} (see below) matched by that of a SSP that has a 
luminosity-weighted metallicity and age that is consistent with values deduced from 
observations of resolved stars. While this is an important check of the 
luminosity-weighted metallicity, it is not a tight constraint on the 
luminosity-weighted age, as very different SFHs can produce similar 
luminosity-weighted ages. Indeed, changing the assumed age by many Gyr results 
in only subtle differences in the spectrum of systems that are older than a few Gyr. 
Still, that the depths of absorption features are more-or-less reproduced indicates 
that (1) nebular continuum emission is likely not a major contributor to the NIR 
SED, as expected from the Byler et al. (2017) models if star formation 
was initiated more than a few Myr in the past, and (2) stochastic sampling 
uncertainties likely do not bias the properties of the F2 spectra. 

\vspace{0.3cm}
\noindent{2)} The NaI$2.21\mu$m and CaI$2.26\mu$m indices vary with location 
in the NSC. The variations are in the sense that the equivalent widths of 
NaI$2.21\mu$m and CaI$2.26\mu$m near the center of the NSC are significantly 
larger than near the ends of the F2 slit. The same results are found 
after light from bright stars has been excised from the spectra, indicating that 
positional changes in the NaI$2.21\mu$m and CaI$2.26\mu$m indices along the 
slit are not restricted to the most luminous -- and hence most evolved -- 
stars near the center of the NSC. 

	Gallego-Cano et al. (2020) examine the spatial distribution 
of bright giants and red clump (RC) stars in the NSC.
While lacking color information to distinguish spectral type, they argue 
that the RC sample is not contaminated by early-type stars, as spectroscopic 
surveys indicate that these tend to be clustered in the central 0.5 parcsec of 
the NSC. Still, the identification of stars based solely on photometry in 
this highly extincted region can be mis-leading (e.g. Nishiyama et al. 2016). 

	Gallego-Cano et al. (2020) find that the ratio 
of bright giants to RC stars varies with distance from the center of the NSC 
outside of the central 0.5 parcsecs, where there has been recent 
star formation, and they interpret this as a consequence of 
incomplete mixing of populations that have different ages -- i.e. there is an 
age gradient. Such a gradient might be expected if the central regions of the NSC 
have been the site of previous star-forming episodes over an extended period of 
time given the finite time required for stars 
to diffuse away from the center of the NSC. That 
the NaI$2.21\mu$m and CaI$2.26\mu$m indices in Regions 1 and 4 differ 
from those in Regions 2 and 3 is consistent with a gradient in the properties of 
cool stars in the NSC. While we lack adequate spectroscopic diagnostics 
to attribute convincingly the gradients in these indices to age, the indices 
measured from the model spectra decrease in strength with increasing age, 
which is {\it qualitatively} consistent with the sense of the expected age gradient 
suggested by Gallego-Cano et al. (2020). However, a gradient solely in age 
can not explain the very large NaI$2.21\mu$m and CaI$2.26\mu$m indices found 
in Regions 2 and 3. 

	The NaI$2.21\mu$m and CaI$2.26\mu$m indices in 
Regions 1 and 4 match those predicted by models that 
have solar abundance mixtures, whereas near the center of the NSC 
they are deeper than predicted by the same models. The NaI$2.21\mu$m and 
CaI$2.26\mu$m indices are not sensitive to uncertainties in plausible 
age and metallicity estimates. Moreover, the deeper than 
expected NaI and CaI indices are found after light 
from the brightest stars is removed. This argues that Na and Ca abundances 
near the center of the NSC are larger than expected for a solar chemical mixture 
over a range of stellar evolutionary states, and not just among the luminous stars 
where deeper than expected Na I and Ca I features were first reported 
by Sellgren et al. (1987) and Blum et al. (1996). More recently, 
Feldmeier-Krause et al. (2017b) also find NaI lines 
in the spectra of stars in the NSC that are deeper than in stellar 
libraries, which presumably reflect solar metallicities and chemical mixtures.

	The depths of Na I lines are sensitive to surface 
gravity, and during the past decade various Na features have been used to examine 
the mass function in integrated light. NaI$2.21\mu$m is 
not as sensitive to the mass function as other Na lines at visible/NIR wavelengths, 
but is an effective probe of [Na/Fe] (Alton et al. 2018). The deep NaI$2.21\mu$m 
indices in Regions 2 and 3 might then suggest that the majority 
of stars in the central regions of the NSC formed from material that originated in 
an environment that produced super-solar abundances of either Na and/or of 
elements and molecules with transitions that are also sampled by the 
NaI$2.21\mu$m index, such as Sc, V, and CN (Carr et al. 2000). A non-solar 
chemical mixture has been proposed for stars near the center of the NSC by Do 
et al. (2018), although those results have been challenged by Thorsbro et al. (2018). 

	Non-solar Na mixtures are not typical of chemical enrichment in a 
quiescent environment like a thin disk. Rather, a Na over-abundance could result in 
material that experienced rapid enrichment by very massive stars (e.g. Kobayashi et 
al. 2006) with suppressed subequent enrichment from lower 
mass stars. A top-heavy IMF will also result in an over-abundance 
of Na. Such abundances are reminiscent of those in globular clusters, for 
which exotic chemical enrichment paths have been proposed (e.g. Denissenkov \& 
Hartwick 2014; Gieles et al. 2018). Massive spheroids show steep gradients in 
[Na/Fe] (Alton et al. 2018; Sarzi et al. 2018), suggesting that the Na yield changes 
with radius in these systems. The behaviour of NaI$2.21\mu$m at locations along 
the F2 slit is reminiscent of that seen in vastly larger systems, albeit on a much 
smaller scale.

	In contrast to the central regions of the NSC, the NaI$2.21\mu$m and 
CaI$2.26\mu$m indices in Regions 1 and 4, where the fractional contribution by light 
from NB and older bulge stars is much higher than near the center of the NSC, 
have equivalent widths that are consistent with a scaled-solar chemical mixture. 
Lee et al. (2018) examine the spectra of red 
clump stars at high latitudes in the Galactic bulge, and find similarities with the 
spectroscopic properties of their counterparts in globular clusters. However, the 
stars observed by Lee et al. (2018) are not in the NB. Some models of bulge 
formation predict radial gradients in stellar content, with stars in the 
NSC and close to the disk plane forming from disk material (e.g. Buck et al. 2018).

\vspace{0.3cm}
\noindent{3)} The NIR emission line characteristics 
of the NSC are consistent with it being a `typical' nucleus 
in terms of the fractional amount of NIR light that comes from recently formed stars. 
This further reinforces the notion that the stellar content 
of the Galactic NSC is broadly similar to nuclear 
clusters in other galaxies (e.g. Neumayer et al. 2020).
When light from the old bulge, the NB, the NSC, and the youngest stars are included, 
the integrated spectrum of the NSC shows (fortuitously!) good agreement 
with the emission and absorption components of the spectrum of the early-type 
Virgo cluster disk galaxy NGC 4491. The agreement between the equivalent 
widths of emission lines does not indicate similar total SFRs. Rather, 
given the vastly different distances of the GC and NGC 4491 then it indicates similar 
densities of ionizing stars and older stars in the volumes sampled by each spectrum. 
The specific SFRs in the areas sampled are thus likely similar.

	The similarity in the depths of the absorption components of the 
SgrA and NGC 4491 spectra reflects the comparable luminosity-weighted ages and 
metallicities of the areas sampled in NGC 4491 and by the F2 slit near the 
GC. Of particular note is that the NaI$2.21\mu$m and CaI$2.26\mu$m features 
in the NGC 4491 and SgrA spectra have very similar 
depths, whereas they are much shallower in the NGC 7793 and NGC 253 spectra. 
This does not mean that the comparatively deep NaI$2.21\mu$m and CaI$2.26\mu$m 
features in the NSC spectrum are abscent near the centers of those 
galaxies, as angular resolution limitations may prevent the 
detection of super-solar NaI and CaI features 
if they are restricted to the compact central regions which may not yet 
have been resolved. 
Indeed, the area over which super-solar NaI$2.21\mu$m and CaI$2.26\mu$m absorption is 
found is very different in NGC 4491 and the MW, occuring over a much larger area 
in the Virgo cluster galaxy than in the central regions of the NSC.

\vspace{0.3cm}
\noindent{4)} An integrated spectrum of the young population in the central regions 
of the NSC has been extracted by subtracting a SSP model spectrum that serves as 
a proxy for the intermediate age and old stars along the NSC sight line. 
The results are not sensitive to the age and metallicity of the model that is 
subtracted from the spectrum. The spectrum of the young component 
that results when light from bright stars is removed is 
largely continuum-dominated, while the spectrum constructed 
with light from bright stars intact shows prominent absorption feaures that are 
attributed to evolved red stars.

	The equivalent widths of many of the emission lines are similar to those 
in the central regions of NGC 253. This does not indicate that the 
innermost regions of the Galaxy is experiencing a wide-scale starburst like that in 
NGC 253, as NGC 253 is much more distant than the GC, and so a larger volume is 
sampled in that galaxy (Section 5). Rather, the similarity in equivalent widths 
indicates that the densities of ionizing sources in the volumes sampled 
are similar. Such agreement was foreshadowed by the similarity between 
excitation conditions in the inner Galaxy and M82 noted by Simpson et al. (1999), 
who described the GC as an aging star burst. It has been suggested that the 
SFR near the GC is low, possibly due to the internal gas kinematics of molecular 
clouds in this area (e.g. Kauffmann et al. 2017). However, the comparison with 
NGC 253 suggests that this is not the case near the center of the NSC.

	That the integrated NIR spectrum of the youngest stars near the GC is 
similar to that of NGC 253 is indirect evidence that the young stars detected 
in the central parsec did not form during a single event, but that star formation has 
occured more-or-less continuously for many Myr near the center of the NSC. 
The basis for this claim is that -- barring a fluke coincidence in SFHs -- 
the similarity between the equivalent widths of emission 
lines in NGC 253 and the GC can plausibly be attributed to the 
convergence of spectroscopic properties in systems that form stars in a continuous 
(or near-continuous) manner, as predicted by models of spectrophotometric evolution 
(e.g. Byler et al. 2017). Barnes et al. (2017) conclude that the SFR in the inner
regions of the Galaxy has been more-or-less continuous during the past few Myr. 
We note that the star-forming activity near the GC need not be strictly continuous to 
produce convergence in emission line properties, and could result if individual star 
formation episodes are separated on Myr timescales (i.e. a large fraction 
of the evolutionary timescales of very massive stars), thereby producing gaps in 
activity that could still maintain a steady supply of ionizing radiation.

	An age dispersion among young stars near the center of the NSC would suggest 
that if star formation has occured more-or-less continuously {\it in situ} then 
either star-forming material can survive for many Myr in the GC environment 
or that there is a steady supply of molecular clouds to replace those that are 
destroyed by tidal forces related to SgrA* (e.g. discussion by Tsuboi et al. 2016).
The distribution of OB stars (e.g. Feldmeier-Krause et al. 2015) supports the notion 
of {\it in situ} star formation near the center of the NSC. The very recent SFH of 
the central regions of the NSC may then not be too different from that in massive 
young clusters in the Galactic disk, despite the obvious differences in environment. 
Evidence for star formation that extends over a few Myr has been found in 
young massive Galactic clusters, such as Westerlund 1 (e.g. Lim et al. 2013), while 
the age range in larger star-forming complexes may exceed 10 Myr (e.g. de Marchi 
et al. 2011). There is also evidence that star formation in clusters may not be 
continuous, but might be distributed over a few episodes (e.g. Beccari et al. 2017). 

	There are obvious avenues for future work. Each of the four regions 
considered in this paper sample $\sim 50 - 65$ arcsec$^2$ along the major axis of the 
NSC, and there is an obvious motivation to obtain integrated light spectra over 
a larger angular area at large angular offsets from the center of the NSC, 
with particular emphasis on the interval 2 -- 4 parsecs along the NSC major axis, 
where star counts indicate that the stellar content 
is changing (Gallego-Cano et al. 2020). The heavy extinction 
at Galactic latitudes south of the GP restricts such work to the 
area north of the GP. IFUs are an obvious tool for expanding sky coverage, 
although there are potential issues with sky subtraction, as a suitably large 
area of high obscuration must be observed to monitor the background in an 
absolute sense. This may bias studies with the best sky subtraction 
to those that sample small fields of view. A merit of long slit spectroscopy 
is that the areas of high obscuration that are close to the GC define 
extended filaments that run more-or-less parallel to the GP, making it 
possible to monitor background light close to the center of the NSC.

	Background light considerations mean that not all existing 
IFU spectroscopic GC datasets may be suitable for 
the study of integrated light. Some have been recorded 
with the goal of examining only individual sources, for which only a local sky 
background must be known (e.g. Do et al. 2013). There are possible exceptions: Eckart 
et al. (2013), Feldmeier-Krause et al. (2015), and Pei$\beta$ker et al. (2020) 
monitored background light by observing a dark cloud $\sim 14$ arcmin from the GC. 
However, in the case of the KMOS spectra obtained by Feldmeier-Krause et al. (2015) 
only one quarter of the IFUs sampled the dark cloud, and so it is not clear if the 
background is known well enough across the mosaiced 
area to obtain reliable integrated light spectra. A limitation of the 
extant IFU spectroscopic surveys is that they cover only the central parsec of the 
NSC, and so do not sample the NSC out to the radii covered by the F2 spectra. 

	Observations with a higher wavelength resolution would also be of interest 
to determine if the NaI$2.21\mu$m and CaI$2.26\mu$m features track 
contributions from Na and Ca, or are skewed by other elements as suggested by Carr 
et al. (2000) in their study of IRS 7. A consideration is that the velocity 
dispersion of the stars in the area studied define inherent limits on 
the wavelength resolution of integrated light spectra. The 
limiting wavelength resolution of integrated light spectra throughout much of the 
area sampled by F2 is $\frac{\lambda}{\Delta \lambda} 
\sim 5000 - 10000$. The observations shown in Figure 6 of Carr 
et al. (2000) suggest that such a spectroscopic resolution should be sufficient 
to resolve lines of Sc and V that are close to the $2.21\mu$m NaI lines. 
There is a rise in the intrinsic velocity dispersion 
near SgrA*, with $\sigma \sim 160$ km/sec at 1 arcsec radius (Trippe et al. 2008). 
The peak in velocity dispersion in the central arcsec 
sets the limiting spectral resolution to a few thousand, complicating 
efforts to resolve these lines in integrated light. 

	Spectroscopic observations with a higher angular resolution will also 
be of obvious interest for probing the central arcsec around SgrA*, as this is 
where the stellar density effects are most extreme, 
and the impact of the severe environment near SgrA*  
most pronounced. One result may be the formation of objects that are 
unique to this environment (e.g. Ciurlo et al. 
2020). There is a dearth of bright late-type stars in the 
central arcsec (e.g. Do et al. 2009), while the spectra of the GC 
in the $2 - 2.4\mu$m wavelength region discussed by Figer et al. (2000) 
show that the light within an arcsec of SgrA* is dominated by young stars. 
Most of the early-type stars in this area have spectral-type B (Eisenhauer et al. 
2005) and are on the main sequence (Habibi et al. 2017). 

	We close by noting that high angular resolution 
observations of integrated light may also allow 
the mass function among low mass stars near SgrA* to be probed with the 
NaI$2.21\mu$m doublet, although there is degeneracy with [Na/Fe]. While the number 
density, spatial distribution, and mass function of sub-solar mass main sequence 
stars have yet to be explored in the central arcsec, it is intriguing 
that the $2.21\mu$m NaI feature in the Figer et al. (2000) 
spectra diminishes in strength with decreasing 
projected distance from SgrA*, although the spectra in this area will have to 
be modelled with an appropriate mix of stellar content. While observations of 
other Na transitions may lift the degeneracy with [Na/Fe], observing 
Na transitions shortward of $\sim 1.3\mu$m in the NSC is 
problematic due to the high line-of-sight extinction. The 
NaI doublet near 2.34$\mu$m may provide useful supplementary information to 
interpret the NaI$2.21\mu$m results, as this feature is prominent in late 
M dwarfs (e.g. Rayner et al. 2009). A complicating factor is that 
this feature is located close to the $^{13}$CO 2--0 band head at 
$2.345\mu$m, compromising detection in spectra with low wavelength resolution. This 
feature is not detected convincingly in the F2 spectra, but may be present 
in the spectra discussed by Feldmeier et al. (2014) in their Figure 3. 

\acknowledgements{It is a pleasure to thank the anonymous referee for 
providing a prompt, comprehensive, and helpful review that greatly improved 
the paper.}

\end{document}